\documentclass[a4paper,11pt]{article}
\usepackage{fullpage}

\usepackage[utf8]{inputenc}
\usepackage[T1]{fontenc}

\usepackage{listings}
\usepackage{xcolor}
\usepackage{xspace}
\usepackage{framed}
\usepackage{lipsum}
\usepackage{amsthm}
\usepackage{amsmath}
\usepackage{hyperref}
\usepackage{float}
\usepackage{placeins}
\usepackage{amsfonts}
\usepackage{array}
\usepackage{booktabs}
\usepackage{tikz}
\usetikzlibrary{positioning, shapes.geometric, trees}
\usepackage{pgfplots}
\pgfplotsset{compat=1.18}

\usepackage{graphicx}
\usepackage{listings}
\usepackage{caption}
\usepackage{subcaption}
\usepackage{enumitem}
\usepackage{algorithm}
\usepackage{algpseudocode}

\definecolor{shadecolor}{gray}{0.75}
\definecolor{codegreen}{rgb}{0,0.6,0}
\definecolor{codegray}{rgb}{0.5,0.5,0.5}
\definecolor{codepurple}{rgb}{0.58,0,0.82}
\definecolor{backcolourcode}{rgb}{0.95,0.95,0.92}

\lstdefinestyle{codestyle}{
    backgroundcolor=\color{backcolourcode},   
    commentstyle=\color{codegreen},
    keywordstyle=\color{magenta},
    numberstyle=\tiny\color{codegray},
    stringstyle=\color{codepurple},
    basicstyle=\ttfamily\scriptsize,
    breakatwhitespace=false,         
    breaklines=true,                 
    captionpos=b,                    
    keepspaces=true,                 
    numbers=left,                    
    numbersep=5pt,                  
    showspaces=false,                
    showstringspaces=false,
    showtabs=false,                  
    tabsize=1,
}

\lstnewenvironment{code}[1][]
  {\lstset{style=codestyle}
  
  \lstset{#1}}
  {}
  
\lstnewenvironment{propertyfile}[1][]
  {\lstset{style=codestyle}
  
  \lstset{#1}}
  {}

\lstnewenvironment{graphfile}[1][]
  {\lstset{style=codestyle}
  
  \lstset{#1}}
  {}

\lstnewenvironment{decompositionfile}[1][]
  {\lstset{style=codestyle}
  
  \lstset{#1}}
  {}  
\lstnewenvironment{instructivedecompositionfile}[1][]
  {\lstset{style=codestyle}
  
  \lstset{#1}}
  {}   
\lstnewenvironment{command}[1][]
  {\lstset{style=codestyle}
  
  \lstset{#1}}
  {}
\lstnewenvironment{outputfile}[1][]
  {\lstset{style=codestyle}
  
  \lstset{#1}}
  {}     

\theoremstyle{plain}
\newtheorem{theorem}{Theorem}[section]

\theoremstyle{definition}
\newtheorem{definition}[theorem]{Definition}

\theoremstyle{remark}

\newcommand{\treewidzard}{\textsc{TreeWidzard}\xspace}

\newcommand{\codeinline}[1]{\texttt{#1}}

\usepackage{listings}
\definecolor{codegreen}{rgb}{0,0.6,0}
\definecolor{codegray}{rgb}{0.5,0.5,0.5}
\definecolor{codepurple}{rgb}{0.58,0,0.82}
\definecolor{grayBackground}{rgb}{0.9,0.9,0.9}
\definecolor{greenBackground}{rgb}{0,0.6,0}
\lstdefinestyle{grayBackground}{
    backgroundcolor=\color{grayBackground},   
    commentstyle=\color{codegreen},
    keywordstyle=\color{magenta},
    numberstyle=\tiny\color{codegray},
    stringstyle=\color{codepurple},
    basicstyle=\ttfamily\footnotesize,
    breakatwhitespace=false,         
    captionpos=b,                    
    keepspaces=true,                 
    numbersep=5pt,                  
    showspaces=false,                
    showstringspaces=false,
    showtabs=false,                  
    tabsize=2
}
\usepackage{listings}
\definecolor{codegreen}{rgb}{0,0.6,0}
\definecolor{codegray}{rgb}{0.5,0.5,0.5}
\definecolor{codepurple}{rgb}{0.58,0,0.82}
\definecolor{grayBackground}{rgb}{0.9,0.9,0.9}
\definecolor{greenBackground}{rgb}{0.8,1,0.8}
\lstdefinestyle{greenBackground}{
    backgroundcolor=\color{greenBackground},   
    commentstyle=\color{codegreen},
    keywordstyle=\color{magenta},
    numberstyle=\tiny\color{codegray},
    stringstyle=\color{codepurple},
    basicstyle=\ttfamily\footnotesize,
    breakatwhitespace=false,         
    captionpos=b,                    
    keepspaces=true,                 
    numbers=left,                    
    numbersep=5pt,                  
    showspaces=false,                
    showstringspaces=false,
    showtabs=false,                  
    tabsize=2
}

\title{\treewidzard: An Engine for Width-Based Dynamic Programming and Automated Theorem Proving}
\date{}

\author{Mateus de Oliveira Oliveira$^{1,2}$ and Sam Urmian$^{2,3}$\\
$^1$Department of Computer and Systems Sciences, Stockholm University, Sweden\\
$^2$Department of Informatics, University of Bergen, Norway\\
$^3$Centre for the Science of Learning \& Technology, University of Bergen, Norway\\
\vspace{0.25em}\\
oliveira@dsv.su.se \hspace{0.5cm} sam.urmian@uib.no}

\begin{document}

\maketitle

\begin{abstract}
In this work, we introduce \treewidzard, an engine for developing dynamic programming
algorithms that decide graph-theoretic properties parameterized by treewidth and pathwidth. Besides
providing a unified framework for algorithms deciding atomic
graph-theoretic properties, our engine allows one to combine such algorithms for two purposes: to obtain dynamic programming algorithms for more complex graph properties, and to support treewidth-based automated theorem proving. Within this context, given the specification of a Boolean combination $P$ of graph properties $P_1, P_2, \ldots, P_r$, and a positive integer $k$, our engine can be used to determine whether all graphs of treewidth at most $k$ satisfy $P$. The main goal of the present work is to provide a system description of \treewidzard. In particular, we provide a step-by-step account of how to implement dynamic programming algorithms in our framework and how to combine these algorithms for model checking and automated theorem proving.
\end{abstract}

\section{Introduction}
{\sloppy

Many NP-hard graph properties, such as chromatic number~\cite{fellows2011complexity}, independent set~\cite{arnborg1989linear}, and Hamiltonian cycle~\cite{bodlaender2015deterministic}, can be decided in fixed-parameter tractable time when parameterized by graph width measures such as treewidth and pathwidth~\cite{arnborg1989linear,bodlaender1997treewidth,bodlaender2007treewidth}. This means that these problems can be solved in time $f(k)\cdot n^{O(1)}$ where $n$ is the number of vertices of the graph and $f$ is a function that depends only on the width parameter $k$. This running time is typically achieved by dynamic programming algorithms that process decompositions from the leaves towards the root.

A separate recurrent need is \emph{reasoning} over bounded-width graph classes: determining whether all graphs of treewidth/pathwidth at most $k$ satisfy a given graph property. The most prominent framework for width-based reasoning on graphs of bounded treewidth/pathwidth uses machinery related to Courcelle's model checking theorem and its automata-theoretic implementations~\cite{courcelle1990monadic,courcelle2000linear,courcelle2012automata,langer2012evaluation}. More specifically, a graph property is specified using a formula $\varphi$ in the monadic second-order logic of graphs, and automata-theoretic techniques are used to compile a tree automaton $\mathcal{A}(\varphi)$ accepting terms encoding models of $\varphi$ of treewidth at most $k$. As a consequence, one can determine whether all graphs of width at most $k$ satisfy $\varphi$ in time $f_{\varphi}(k)$, for some function $f_\varphi$.

The main drawback of the Courcelle-based approach for width-based automated theorem proving is that the function $f_{\varphi}(k)$ grows as a tower of exponentials whose height is the number of quantifier alternations in $\varphi$~\cite{frick2004complexity}. As a consequence, even if the formula defining a given graph property has few alternations, the running times of the resulting algorithms are prohibitively large, even for small width.

A recent line of work~\cite{de2023width,de2024state} introduced an algorithmic approach for width-based automated theorem proving where graph properties are specified using dynamic programming algorithms instead of logical specifications. In this framework, the complexity of determining whether all graphs satisfy a given property $P$ is double exponential in the size of the local information used by the dynamic programming algorithms. This approach has been used to show that several prominent conjectures in graph theory, such as Hadwiger's conjecture and Tutte's flow conjectures, among many others, can be tested in time double exponential in $k$ on graphs of treewidth at most $k$~\cite{de2023width}. Additionally, symmetry breaking and early-pruning techniques developed in~\cite{de2024state} allow the approach to be evaluated in practice. More specifically,~\cite{de2024state} confirmed the validity of Reed's triangle-free conjecture on graphs of pathwidth at most $5$ and of treewidth at most $3$.

\subsection{Our Contributions}

In this work, we provide a \emph{system description} of \treewidzard, a C++ engine for specifying dynamic programming algorithms (DP-cores) that operate on tree and path decompositions. The implementation is publicly available~\cite{treewidzard}. Our engine implements and extends the core ideas from the width-based automated theorem proving framework developed in \cite{de2023width,de2024state}. Dynamic programming algorithms developed in \treewidzard serve two main purposes: first, computing graph invariants and deciding graph properties. Second, enabling width-based automated theorem proving. In this setting, given a Boolean combination $P$ of graph properties $P_1, P_2, \ldots, P_r$ and a positive integer $k$, the engine can determine whether all graphs of treewidth/pathwidth at most $k$ satisfy $P$. If this is not the case, and the DP-cores defining $P_1,\ldots,P_r$ satisfy the finiteness and coherence conditions discussed below, a counterexample graph of width at most $k$ that does not satisfy $P$ is produced.

Our tool supports the processing of tree decompositions in two formats. Natively, graphs of treewidth at most $k$ and their associated tree decompositions are specified algebraically in \treewidzard using the notion of a $k$-instructive tree decomposition (ITD). Intuitively, such
a decomposition is a term that specifies the construction of a graph of treewidth at most $k$. In the second format, graphs and tree decompositions are specified in a more traditional way, using two structures: one for the graph and another for the decomposition. More specifically, this is the DIMACS/PACE format used in the 2017 edition of the PACE challenge~\cite{dell2018pace}. Decompositions in this format are automatically converted into ITDs for processing by \treewidzard.

Our primary focus is modularity. In particular, DP-cores developed in \treewidzard are designed as independent components that can be combined to define more complex DP-cores. Additionally, for each fixed $k$, the same engine used to process individual $k$-instructive tree decompositions for model checking and invariant computation can also be used for automated theorem proving: that is, determining whether all graphs of treewidth at most $k$ satisfy a given graph property. If a given statement expressible as a Boolean combination of graph properties defined by DP-cores is not valid on the class of graphs of treewidth/pathwidth at most $k$, then \treewidzard is guaranteed to produce a concrete counterexample, provided the DP-cores involved in the statement satisfy certain finiteness and coherence properties.

From an optimization perspective, \treewidzard incorporates the techniques introduced in~\cite{de2024state} for state canonization (symmetry breaking by state isomorphism) and premise-based pruning for subgraph-closed premises. These optimizations significantly reduce the space of explored states while preserving soundness and refutational completeness for width-based automated theorem proving. In this work, we provide a detailed account of how to use \treewidzard for model checking, invariant computation, and width-based automated theorem proving. We also provide a tutorial-like exposition on how to implement DP-cores in \treewidzard.

\subsection{Related Work}\label{sec:related-work}

Courcelle-style meta-theorems are commonly presented as MSO model checking on bounded treewidth via
compilation of a formula into an automaton/game evaluated over a decomposition~\cite{courcelle1990monadic,seese1991structure}.
The same automata-theoretic route can, in principle, be repurposed for width-bounded automated theorem
proving: to establish $\varphi \Rightarrow \psi$ over all width-$k$ graphs, one can compile to decomposition
automata and reduce the claim to a language-inclusion check $L(\varphi)\subseteq L(\psi)$ (equivalently,
emptiness of $L(\varphi\wedge \neg\psi)$). In practice, however,
the dependence on formula structure can be severe: alternation depth and compositional growth can cause
large intermediate automata and substantial constants even when width is fixed, and inclusion/complement
steps may further amplify these effects~\cite{courcelle1990monadic}.
\treewidzard instead searches the induced global execution-state space of DP-cores over ITDs,
enabling incremental composition and leveraging canonization/pruning to manage blowup. Such optimization steps are inherent in our dynamic programming view and do not have a counterpart in the purely automata-theoretic formalism.

A complementary line of research studies decomposition-first dynamic programming frameworks and declarative systems
that exploit supplied decompositions, often achieving strong instance-level performance for
satisfaction/optimization tasks~\cite{fichte2017answer,bannach2023structure}. Related declarative toolchains
(e.g., ASP systems such as clingo) provide extensible backends that can host decomposition-based methods
in practice~\cite{gebser2016theory}. 
Additionally, first-order (FO) model checking has well-studied tractability frontiers tied to locality and structural restrictions, and in practice it is frequently supported by SAT/SMT and first-order ATP backends~\cite{de2008z3,barrett2011cvc4,kovacs2013first,weidenbach2009spass}. 
Nevertheless, the primary focus of these approaches remains per-instance evaluation rather than the evaluation of a given property on a whole class of graphs of bounded width. 

\treewidzard bridges Courcelle-style compilation and custom decomposition-based dynamic programming by exposing dynamic programs as composable DP-cores over a fixed ITD instruction set~\cite{de2023width,de2024state}. In this way, it provides a formalism that reconciles the automata-based approach, in which tree automata for graph classes are typically compiled from logical specifications, with the dynamic-programming approach: in our setting, these automata are compiled instead from dynamic programming algorithms. This framework can be used both to check whether a given graph satisfies a property and to determine whether every graph of treewidth or pathwidth at most $k$ satisfies that property. When the latter does not hold, \treewidzard produces a counterexample.

}

\section{Preliminaries}
\label{sec:preliminaries}

Treewidth and pathwidth are used in their standard sense: $\operatorname{tw}(G)\le k$ means that
$G$ has a tree decomposition with bag size at most $k{+}1$, and
$\operatorname{pw}(G)\le k$ is the corresponding path-decomposition variant.
\treewidzard does not manipulate bag-labeled trees directly. Instead, it uses an instruction-term
representation (ITD) that encodes both the constructed graph and the decomposition shape.

Fix $k\in\mathbb{N}$ and let $[k{+}1]=\{1,\dots,k{+}1\}$ be a pool of labels.
At each node we keep an active-label set $B\subseteq [k{+}1]$, which plays the role of the current bag,
together with a partial map $\ell:[k{+}1]\rightharpoonup V_{\mathrm{curr}}$ whose domain is exactly $B$,
where $V_{\mathrm{curr}}$ is the set of vertices created so far at that node.
The corresponding vertex bag is $\ell(B)$.
Each active label names exactly one current vertex, and distinct active labels denote distinct
current vertices.
Labels may be forgotten and reused, so the width bound controls the number of simultaneously
active labels, not the total number of created vertices. Forgetting a label removes it from the
active set, not from the graph. Each $\mathsf{IntroVertex}_u$ creates a fresh graph vertex;
reusing a forgotten label creates a new vertex, not the previous one.
In this paper, ``active labels'' always refers to these temporary slots in $[k{+}1]$; external
file formats use separate graph vertex IDs.

\begin{figure}[t]
  \centering
  \includegraphics[scale=0.8]{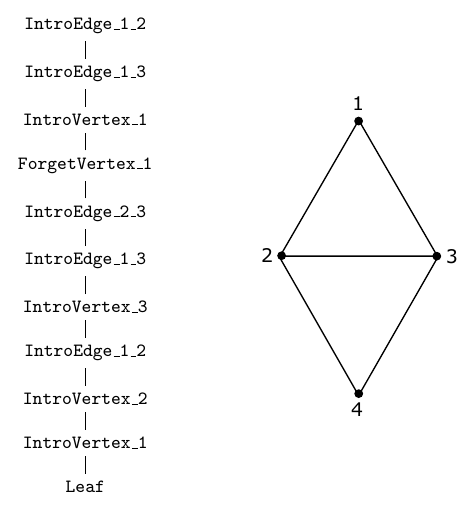}
  \caption{A $2$-instructive path decomposition (with \textsf{Leaf} at the bottom;
  evaluation is bottom-up) and the graph it constructs. Vertex names in the graph indicate
  creation order, not label identities. Forgetting a label releases it for later reuse, so the
  term may create more than $k{+}1$ vertices while never having more than $k{+}1$ active labels
  at any node.}
  \label{fig:itd-example}
\end{figure}

Figure~\ref{fig:itd-example} illustrates this reuse on a width-$2$ term constructing the diamond graph.

\begin{definition}[ITD instruction set and terms]
\label{def:itds}
Fix $k\in\mathbb{N}$. The instruction alphabet is
\[
\Sigma_k=
\{\mathsf{Leaf},\mathsf{Join}\}
\cup\{\mathsf{IntroVertex}_u,\mathsf{ForgetVertex}_u : u\in [k{+}1]\}
\]
\[
\cup\{\mathsf{IntroEdge}_{u,v} : u,v\in [k{+}1],\ u\neq v\}.
\]

A rooted instruction tree over $\Sigma_k$ is a \emph{$k$-instructive tree decomposition} if each node
$t$ has an active-label set $B(t)\subseteq [k{+}1]$ and all local side conditions hold:
$\mathsf{IntroVertex}_u$ introduces only inactive labels, $\mathsf{ForgetVertex}_u$ forgets only active labels,
$\mathsf{IntroEdge}_{u,v}$ is allowed only when $u$ and $v$ are active and distinct, and
$\mathsf{Join}$ combines children with equal
active-label sets.
These active-label sets are computed inductively and are uniquely determined:
$B(\mathsf{Leaf})=\emptyset$;
$\mathsf{IntroVertex}_u$ adds label $u$; $\mathsf{ForgetVertex}_u$ removes label $u$;
$\mathsf{IntroEdge}_{u,v}$ preserves the active-label set;
and $\mathsf{Join}$ requires the two child sets to be equal.

We write $\mathrm{ITD}_k$ for the set of such instruction trees and $\mathrm{IPD}_k\subseteq\mathrm{ITD}_k$
for the join-free sublanguage (instructive path decompositions). Join-free terms are unary chains,
corresponding to path decompositions.
Terms may end with a nonempty active-label set; appending trailing $\mathsf{ForgetVertex}$ steps yields
a term that constructs the same graph (up to $\cong$) and does not increase width, with an empty
active-label set at the root.
\end{definition}

Each term $\tau\in\mathrm{ITD}_k$ defines a graph $G(\tau)$ by bottom-up evaluation.
Operationally, each node maintains a partial graph together with the current active-label map.
$\mathsf{Join}$ combines two subcomputations with the same active-label set (shared bag/interface):
it takes the disjoint union of child graphs, preserves edges from both branches, and identifies
vertices with equal active labels
(so parallel edges may appear when both branches contribute the same endpoint pair).
After $\mathsf{Join}$, each active label refers to the merged vertex obtained from the two equally
labeled child vertices.

This ITD/IPD view is just a normal form for standard width notions:
\begin{theorem}[ITD/IPD width equals treewidth/pathwidth]
\label{thm:itd-ipd-width}
For every graph $G$,
\[
\operatorname{tw}(G)\le k \iff \exists \tau\in\mathrm{ITD}_k\text{ with }G(\tau)\cong G,
\]
and similarly,
\[
\operatorname{pw}(G)\le k \iff \exists \tau\in\mathrm{IPD}_k\text{ with }G(\tau)\cong G.
\]
\end{theorem}
This equivalence is standard~\cite{de2023width,de2024state}.
Here $\cong$ denotes isomorphism of underlying simple graphs (after suppressing parallel-edge
multiplicities). The internal semantics allows multiedges (mainly due to $\mathsf{Join}$); built-in
cores used in this paper target simple-graph properties and treat multiplicity as irrelevant.

A dynamic program in \treewidzard is specified as a DP-core: a witness representation, one local
transition routine per ITD instruction, and a root decision/invariant routine.
The kernel evaluates these routines bottom-up on an input ITD.
For user-implemented DP-cores (built-in cores satisfy this), the key correctness condition is
isomorphism invariance (decomposition-independence):
acceptance and invariant values should depend only on the underlying graph, not on which equivalent
ITD representation is used, i.e., if $G(\tau_1)\cong G(\tau_2)$ then the core returns the same
acceptance decision and invariant value. The full interface is detailed in
Section~\ref{sec:chromatic-number-dp-core}.

\section{Usage}\label{sec:usage}
\treewidzard evaluates compositions of DP-cores over bounded-width decompositions.
Given a specification (a composition of decision properties and numeric invariants), it can
(i) evaluate the specification on a concrete instance supplied with a decomposition/ITD,
(ii) compute invariants for optimization-style cores on such instances, and
(iii) decide bounded-width validity by ranging over all width-$\le k$ ITDs and searching for a
counterexample.

\begin{figure}[H]
    \centering
    \includegraphics[scale=0.9]{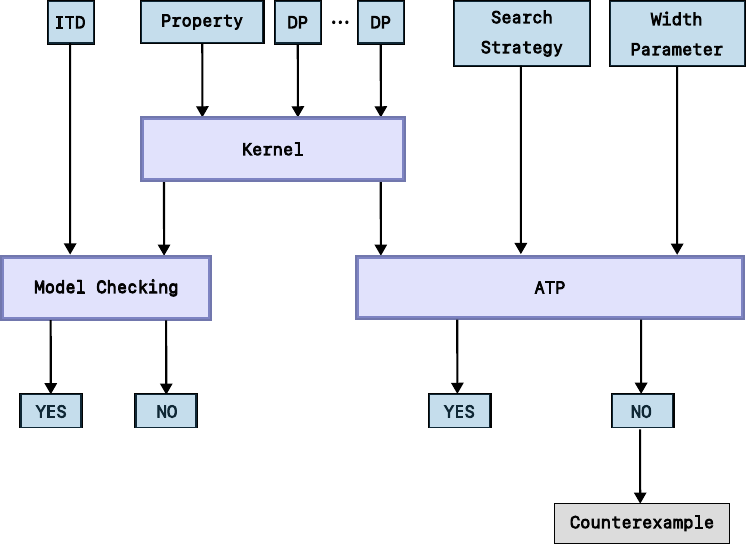}
    \caption{Relation between \treewidzard's classes.}
    \label{fig:class-diagram}
\end{figure}

Figure~\ref{fig:class-diagram} summarizes the separation of concerns in the implementation.
The kernel executes (possibly composed) DP-cores bottom-up on ITDs; the model-checking path
supplies a concrete ITD/decomposition and reports acceptance; and the ATP path drives the
bounded-width exploration over all $k$-instructive tree decompositions, returning either a counterexample graph
or a successful bounded-validity verdict.
The output snippets below omit only the common DP-core loading banner that $\treewidzard$ prints before each run.

\subsection{The Property File}\label{usage:property-file}
In $\treewidzard$, a property file assigns variables to DP-cores and then specifies a formula over these variables and, when appropriate, over the invariant values computed by the cores. A variable itself is Boolean: for a variable \verb|x|, the truth value of \verb|x| records whether the input graph is accepted by the DP-core assigned to \verb|x|. The expression \verb|INV(x)| denotes the numeric value reported by that core. For Boolean DP-cores, \verb|INV(x)| is the same Boolean value as \verb|x|, encoded as $0$ or $1$; for invariant-computing DP-cores it is the actual invariant value.
\begin{code}[]
x := ChromaticNumber_AtMost(3)
y := IndependenceNumber()
Formula
x AND (INV(y) >= 2)
\end{code}

In this example, \verb|x| is true on a graph if and only if the graph is $3$-colorable. The variable \verb|y| is also a Boolean acceptance variable, while \verb|INV(y)| is the independence number (the size of a maximum independent set) computed by the corresponding DP-core. A graph satisfies the property specified by the formula if it is $3$-colorable and has an independent set of size at least $2$. The full syntax for the specification of a property file is described in Appendix \ref{appendix:FormatPropertyFile}.

\subsection{Model Checking}\label{usage:model-checking}

When using $\treewidzard$ for model checking, the goal is to verify whether a given graph satisfies a given graph property. 
The graph property is specified in a property file, while the graph itself can be specified in two ways.
In the first way, we need to provide two separate files—one specifying the edge list of the graph and another specifying a tree decomposition of the graph.
In the second way, we need to provide a single file defining an instructive tree decomposition (ITD), which implicitly encodes both the graph and its tree decomposition. 
The program returns \codeinline{Property Satisfied} if the
graph satisfies the property in question, and \codeinline{Property Not Satisfied} otherwise. 

\subsubsection{Model Checking Given a Graph and a Tree Decomposition}
\label{usage:model checking option one}
\begin{figure}[H]
\centering
\includegraphics[scale=0.8]{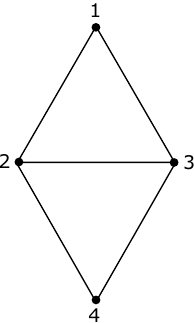}
\caption{The diamond graph used in the model-checking example.}
\label{fig:diamond-graph}
\end{figure}

In this approach, the input consists of a property file and separate files specifying a graph and one of its tree decompositions.
For example, suppose that our goal is to check whether the diamond graph depicted in Figure \ref{fig:diamond-graph} is $3$-colorable. We start by creating a property file, \verb|three_colorable.txt|, defining a formula with a single variable. This formula is true on a graph if and only if the graph is $3$-colorable, meaning that its chromatic number is at most $3$. 
\begin{propertyfile}
x := ChromaticNumber_AtMost(3)
Formula
x
\end{propertyfile}

Next, we create a file named \verb|diamond.gr|, which provides the description of the graph shown in Figure \ref{fig:diamond-graph}. 
Graphs are specified using a variant of the DIMACS format. This variant is the same one used in the 2017 edition of the PACE challenge \cite{dell2018pace}. A detailed description of this format is provided in Appendix \ref{appendix:graph_file}.
In the example below, the first line, starting with \texttt{c}, is a comment line. The second line, starting with \verb|p tw|, has the form \verb|p tw n m|, where \verb|n| is the number of vertices and \verb|m| is the number of edges. In this example, the graph has four vertices and five edges.
Each subsequent line describes a single undirected edge using two integers, each representing a vertex label. Vertices are labeled consecutively from 1 to \texttt{n}. Multiple edges are permitted (self-loops are not supported), as detailed in the appendix. In our example, the edges are $\{1,2\}, \{1,3\}, \{2,3\}, \{2,4\}$, and $\{3,4\}$ (see Figure \ref{fig:diamond-graph}). 
\begin{graphfile}[label={graphfile: diamond}]
c The diamond graph.
p tw 4 5
1 2
1 3
2 3
2 4
3 4
\end{graphfile}

The tree decomposition file, named \verb|diamond.td|, specifies a tree decomposition of the graph, following the PACE 2017 format~\cite{dell2018pace} (see Appendix \ref{appendix:decomposition_file}).
In the example below, the first line, starting with \texttt{c}, is a comment line. The line starting with \texttt{s td} has the format \texttt{s td N w n}, where \texttt{N} is the number of bags, \texttt{w} is the size of the largest bag, and \texttt{n} is the number of vertices in the original graph. In this example, the decomposition consists of 2 bags, the largest bag has size 3, and the graph contains 4 vertices. Consequently, the width of the decomposition is $3 - 1 = 2$, that is, the size of the largest bag minus one.
Each subsequent line, starting with a \texttt{b}, defines a bag. The first number after \texttt{b} is the bag identifier, followed by a list of vertices contained in that bag. In our example, the first bag \verb|b 1| contains vertices 1, 2, and 3, and the second bag \verb|b 2| contains vertices 2, 3, and 4.
Finally, each remaining line, consisting of a pair of numbers, describes an edge connecting two bags of the tree decomposition. In our example, the tree decomposition has a single edge, which connects bag \verb|b 1| to bag \verb|b 2|.

\begin{decompositionfile}[label={decompositionfile: diamond}]
c A tree decomposition of the diamond graph.
s td 2 3 4
b 1 1 2 3 
b 2 2 3 4
1 2    
\end{decompositionfile}

To verify whether the diamond graph is $3$-colorable, we call $\treewidzard$ with the option \verb|-modelcheck PACE| and pass as input the property file (\verb|three_colorable.txt|), the graph file (\verb|diamond.gr|), and the tree decomposition file (\verb|diamond.td|). The order of the files is relevant.

\begin{command}[label={command: testing 3 colorability with pace}]
./treewidzard.sh -modelcheck PACE three_colorable.txt diamond.gr diamond.td
\end{command}

After executing this command, $\treewidzard$ produces the output:
\begin{outputfile}
----------------------------------------------------------
Formula:x

Result:PROPERTY SATISFIED

Execution information:

x := ChromaticNumber_AtMost(3)
Core type: Bool
Final value: x=1
Invariant value: INV(x)=1
Max witness set: 3
-------------------------
\end{outputfile}

The output indicates that the property defined by the file \verb|three_colorable.txt| is satisfied. In addition to this result, the output includes information about the DP-core and the value of the variable \verb|x|. This variable is set to $1$ if the graph is $3$-colorable, and $0$ otherwise. The output further provides the value \verb|INV(x)|, which represents the invariant computed by the DP-core \verb|ChromaticNumber_AtMost|. For this particular DP-core, the invariant is trivial: it takes the value $1$ if the DP-core accepts the graph and $0$ otherwise. Finally, the output includes the size of the largest witness set generated during the evaluation of the property on the input graph.
In this run, \verb|Max witness set: 3| is small because coloring witnesses are partitions of the current bag into color classes, and even a single edge constraint prunes most partitions. Concretely, when the bag is $\{1,2,3\}$ and the edge $\{1,2\}$ is present, the only admissible partitions are $\{\{1\},\{2\},\{3\}\}$, $\{\{1,3\},\{2\}\}$, and $\{\{1\},\{2,3\}\}$.

\subsubsection{Model Checking Given an Instructive Tree Decomposition}\label{usage:model checking option two}

Alternatively, instead of providing separate graph and tree decomposition files, 
we can provide a single file defining an instructive tree decomposition (ITD). 
This file encodes both a graph and a tree decomposition of the graph.
The file named \verb|diamond.itd| specifies a 2-ITD for the graph shown in Figure \ref{fig:diamond-graph}. A detailed description of the ITD file format is provided in Appendix \ref{appendix:itd_file}.

\begin{instructivedecompositionfile}[label={itdfile: diamond}]
1 Leaf
2 IntroVertex_1(1)
3 IntroVertex_2(2)
4 IntroEdge_1_2(3)
5 IntroVertex_3(4)
6 IntroEdge_1_3(5)
7 IntroEdge_2_3(6)
8 ForgetVertex_1(7)
9 IntroVertex_1(8)
10 IntroEdge_1_3(9)
11 IntroEdge_1_2(10)    
\end{instructivedecompositionfile}    

To check whether the diamond graph is $3$-colorable, we call $\treewidzard$ with the parameter \verb|-modelcheck ITD|, passing the property file (\verb|three_colorable.txt|) and the instructive tree decomposition file (\verb|diamond.itd|) as input. The order of the files is relevant.

\begin{command}[label={command: testing 3 colorability with abstract}]
./treewidzard.sh -modelcheck ITD three_colorable.txt diamond.itd
\end{command}

As in the previous case, after executing this command, $\treewidzard$ produces the following output:

\begin{outputfile}
----------------------------------------------------------
Formula:x

Result:PROPERTY SATISFIED

Execution information:

x := ChromaticNumber_AtMost(3)
Core type: Bool
Final value: x=1
Invariant value: INV(x)=1
Max witness set: 3
-------------------------
\end{outputfile}

As above, the witness-set maximum reflects the fact that witnesses are bag partitions: at the DP stage where the bag contains three vertices but only one edge is present, exactly three partitions remain admissible (the two endpoints must be in different blocks).

\subsection{Computing Invariants}\label{usage:computing invariant}

In the previous examples, we used $\treewidzard$ to decide whether a given graph satisfies a given graph property. We now show how to use $\treewidzard$ to compute graph invariants. In this example, our goal is to compute the independence number of a graph, that is, the maximum size of an independent set. We start by creating a file called \verb|independence_number.txt| with the following content:

\begin{code}
x := IndependenceNumber()
Formula
INV(x) == 2
\end{code}

The DP-core \verb|IndependenceNumber| accepts every graph. Therefore, the variable \verb|x| is always true, while the formula tests the numeric value computed by the core. The distinction between \verb|x| and \verb|INV(x)| is important here: \verb|x| is the Boolean acceptance value, while \verb|INV(x)| is the nontrivial graph invariant computed by the core, namely the independence number of the input graph. For Boolean DP-cores these two values coincide, up to the $0$/$1$ encoding of truth values.

To compute the value of this invariant in the diamond graph, depicted in Figure \ref{fig:diamond-graph}, we run the same command used for model checking. We have two possibilities, depending on how the input graph is represented.

\begin{code}  
./treewidzard.sh -modelcheck PACE independence_number.txt diamond.gr diamond.td

./treewidzard.sh -modelcheck ITD independence_number.txt diamond.itd
\end{code}

In either case, $\treewidzard$ produces the following output.

\begin{outputfile}
----------------------------------------------------------
Formula:( INV(x) == 2)

Result:PROPERTY SATISFIED

Execution information:

x := IndependenceNumber()
Core type: Max
Final value: x=1
Invariant value: INV(x)=2
Max witness set: 6
-------------------------
\end{outputfile}

The variable $x$ is equal to $1$, indicating that the graph is accepted by the DP-core, while \verb|INV(x)| is the independence number of the diamond graph, which is $2$. Since the core type is \texttt{Max}, at the root \treewidzard takes the maximum \verb|INV| value across final witnesses. The formula \verb|INV(x) == 2| is therefore satisfied.

\subsection{Automated Theorem Proving}\label{usage:automated theorem proving}

Width-based automated theorem proving is a framework in which the goal is to determine whether all graphs with width at most $k$ (here, treewidth or pathwidth) satisfy a given property~\cite{de2023width}. 

When running ATP, \treewidzard searches for a bounded-width counterexample to the specified property.
For refutations, use \texttt{-files} to export a concrete counterexample that can be rechecked in model-checking mode.
The examples below use \texttt{-pl}, which is the engine flag for printing loop progress, and \texttt{-nthreads 10}, which sets the thread count for parallel search.

For example, suppose we want to test whether every graph with treewidth at most $4$ is $5$-colorable. The first step is to create a property file specifying a formula that is true on a graph if and only if it is $5$-colorable. Let us call this file \verb|five_colorable.txt|.

\begin{propertyfile}[label={propertyfile: 5 colorability}]
x := ChromaticNumber_AtMost(5)
Formula
x
\end{propertyfile}

Next, we need to specify the value of the width parameter (\verb|tw=4|) and the search strategy that will traverse the state space. The example below uses a breadth-first search executed in parallel with 10 threads.
\begin{command}[label={command: atp bfs}]
./treewidzard.sh -atp tw=4 -pl -nthreads 10 ParallelBreadthFirstSearch five_colorable.txt
\end{command}
As a result, $\treewidzard$ produces the following output:

\begin{outputfile}
----------------------------------------------------------
Property information:

Formula: x

x := ChromaticNumber_AtMost(5)
Core type: Bool

Search information:

Width parameter: tree_width = 4
Search method: ParallelBreadthFirstSearch
Premise flag: NOT ACTIVATED

Search process:
Iteration                ALLSTATES                NEWSTATES                Max WITNESSSET SIZE
1                        6                        5                        1
2                        16                       10                       2
3                        36                       20                       5
4                        71                       35                       15
5                        132                      61                       52
6                        227                      95                       52
7                        447                      220                      52
8                        812                      365                      52
9                        1439                     627                      52
10                       1450                     11                       52
11                       1450                     0                        52

Result: PROPERTY SATISFIED
-----------------------------------------
\end{outputfile}

The output shows that all graphs with treewidth at most $4$ can be colored with at most $5$ colors.
Additionally, the program generates $1450$ states, each of which consists of a set of local witnesses~\cite{de2023width}. In this example, the maximum size of this set was $52$.
This maximum is not accidental: since $k=4$ implies bags of size at most $k{+}1=5$, and witnesses are partitions of the bag into color classes, the theoretical ceiling for \texttt{ChromaticNumber\_AtMost(5)} is the Bell number $B_5=52$ (all set partitions of a 5-element bag).

We now consider the question of whether all graphs of treewidth at most $4$ are $4$-colorable.
The following file, called \verb|four_colorable.txt|, defines a formula that is true on a graph if and only if the graph is $4$-colorable.

\begin{propertyfile}[]
x := ChromaticNumber_AtMost(4)
Formula
x
\end{propertyfile}

By running the following command:

\begin{code}
./treewidzard.sh -atp tw=4 -pl -nthreads 10 ParallelBreadthFirstSearch four_colorable.txt
\end{code}
$\treewidzard$ produces the following output, describing a counterexample for the statement that all graphs of treewidth at most $4$ are $4$-colorable.
\begin{outputfile}
----------------------------------------------------------
Property information:

Formula: x

x := ChromaticNumber_AtMost(4)
Core type: Bool

Search information:

Width parameter: tree_width = 4
Search method: ParallelBreadthFirstSearch
Premise flag: NOT ACTIVATED

Search process:
Iteration                ALLSTATES                NEWSTATES                Max WITNESSSET SIZE
1                        6                        5                        1
2                        16                       10                       2
3                        36                       20                       5
4                        71                       35                       15
5                        132                      61                       51
6                        227                      95                       51
7                        447                      220                      51
8                        812                      365                      51
9                        1444                     632                      51

Result: PROPERTY NOT SATISFIED

Counterexample found:
The assignment that makes the formula false:
x=0
-----------------------------------------
\end{outputfile}

\begin{figure}[!h]
\centering
\includegraphics[scale=0.8]{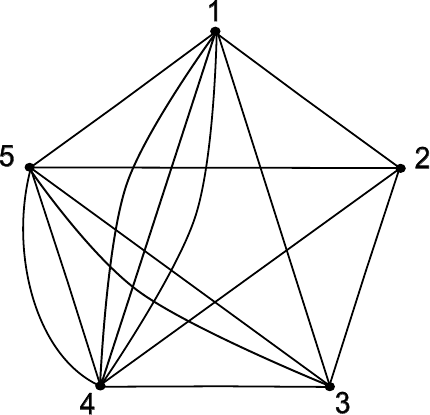}
\caption{
A graph of treewidth $4$ that is not $4$-colorable. This counterexample was generated by $\treewidzard$ while evaluating whether all graphs of treewidth at most $4$ are $4$-colorable.}
\label{fig:counterexample}
\end{figure}

The counterexample was found after analyzing $1444$ states. The output reports the falsifying assignment, namely that the variable \verb|x| is false. If the same run is executed with the \texttt{-files} flag, $\treewidzard$ writes the corresponding counterexample graph and decomposition files. One exported counterexample is depicted in Figure~\ref{fig:counterexample}; its underlying simple graph is $K_5$, which is not $4$-colorable.
Appendix~\ref{appendix:additional_counterexamples} records two further counterexamples generated by $\treewidzard$ for degree-bound sanity checks.

\subsubsection{Symmetry Breaking}
In \treewidzard, the search strategy \texttt{ParallelIsomorphismBreadthFirstSearch} implements symmetry breaking via the state-canonization technique of~\cite{de2024state}. As an example, we apply this search strategy to test whether all graphs of treewidth at most $4$ are $5$-colorable.

\begin{code}
./treewidzard.sh -atp tw=4 -pl -nthreads 10 ParallelIsomorphismBreadthFirstSearch five_colorable.txt
\end{code}

\begin{outputfile}
----------------------------------------------------------
Property information:

Formula: x

x := ChromaticNumber_AtMost(5)
Core type: Bool

Search information:

Width parameter: tree_width = 4
Search method: ParallelIsomorphismBreadthFirstSearch
Premise flag: NOT ACTIVATED

Search process:
Iteration                ALLSTATES                NEWSTATES                Max WITNESSSET SIZE
1                        2                        1                        1
2                        3                        1                        2
3                        5                        2                        5
4                        7                        2                        15
5                        10                       3                        52
6                        14                       4                        52
7                        21                       7                        52
8                        33                       12                       52
9                        51                       18                       52
10                       53                       2                        52
11                       53                       0                        52

Result: PROPERTY SATISFIED

-----------------------------------------
\end{outputfile}

With symmetry breaking, $\treewidzard$ generates 53 states in canonical form.
The earlier breadth-first run generated 1450 states. Here, only canonical states are stored, reducing the total count to 53 and illustrating the effectiveness of symmetry breaking.

\subsubsection{Automated Theorem Proving with Premise Pruning}
\label{subsection:PremiseSearch}

In $\treewidzard$, \emph{premise pruning} can be activated by adding the \codeinline{-premise} flag before the search strategy. This option is useful for implications of the form $A \Rightarrow B$ where the premise $A$ is closed under subgraphs. In this case, once a partial graph violates $A$, every supergraph also violates $A$, so the implication is already true downstream and the search can prune that branch. Premise pruning should be enabled only under this subgraph-closedness condition; otherwise, completeness may be lost.

We illustrate premise pruning using the invariant-returning DP-core \texttt{VertexCount()}, which computes $|V(G)|$. Consider the following property file:
\begin{code}
z := VertexCount()
Formula
( INV(z) <= 20 ) IMPLIES (1 == 1)
\end{code}
The premise $(|V(G)|\le 20)$ is subgraph-closed, so \texttt{-premise} prunes branches as soon as the vertex bound is exceeded. In practice, this turns an otherwise unbounded state space into a finite exploration.
Appendix~\ref{appendix:unbounded_atp} expands on this example, explains why \texttt{VertexCount()} induces an unbounded ATP search space, and shows how premise pruning recovers a finite search on the premise-bounded domain.

\section{Example DP-Core: \texttt{ChromaticNumber\_AtMost}}
\label{sec:chromatic-number-dp-core}
\sloppy
When implementing a DP-core that decides a specific graph property, two key structures must be defined.
The first structure specifies a local witness for the DP-core. Intuitively, such a local witness stores partial information used by the DP-core to decide whether the input graph satisfies the property in question as the input tree decomposition is traversed from the leaves towards the root. The second structure defines the transition functions of the DP-core itself, along with other auxiliary functions.
Each of these structures is discussed in the following subsections, using the DP-core \verb|ChromaticNumber_AtMost| as a concrete example. This DP-core decides whether a graph has chromatic number at most $c$ for a given number $c$ passed as a parameter.

\subsection{Local Witness}
To define a local witness, we need to create a structure that includes the attributes necessary for storing information about the witness. Additionally, we need to implement auxiliary functions, such as those for comparison, relabeling, hashing, and printing. The following is the general format for such a structure.

\begin{code}[language=C++,label={code:witness}
,caption={General structure of a local witness in $\treewidzard$},
]
struct DP_Core_Name_Witness: WitnessWrapper<DP_Core_Name_Witness> {
friend bool is_equal_implementation(const WitnessAlias &l, const WitnessAlias &r){};

friend bool is_less_implementation(const WitnessAlias &l, const WitnessAlias &r){};

WitnessAlias relabel_implementation(const std::map<unsigned, unsigned> &relabelingMap) const{};

void hash(Hasher &h) const override{};

void witness_info(std::ostream &os) const{};
};
\end{code}

Next, we demonstrate how to instantiate the components of such a structure using local witnesses of the DP-core \verb|ChromaticNumber_AtMost| as an example.

\paragraph{Attributes.} The central attribute of this structure is \texttt{partialColoring}, which is a vector of sets of numbers. Intuitively, \verb|partialColoring| is a partition of the active labels associated with the node that is currently being processed in the tree decomposition. Each cell of the partition represents a color class. Two vertices corresponding to active labels share the same color if and only if their labels belong to the same cell of the partition. The implementation keeps the vector sorted so that comparison and hashing use a canonical representation.

\begin{code}[language=C++]
struct ChromaticNumber_AtMost_Witness : WitnessWrapper<ChromaticNumber_AtMost_Witness> {
    std::vector<std::set<unsigned>> partialColoring;
};
\end{code}

\paragraph{Comparison Functions.} Two friend functions are provided to enable the comparison of witness objects. The function \texttt{is\_equal\_implementation} compares two witness objects for equality by directly comparing their \texttt{partialColoring} attributes. In contrast, the \texttt{is\_less\_implementation} function defines a strict total order between two witnesses using the lexicographical order of the canonical vector representation. These comparisons are essential when witness objects are stored in ordered data structures, and when the dynamic programming algorithm needs to make consistent decisions about the order in which witnesses are processed.

\begin{code}[language=C++]
friend bool is_equal_implementation(const WitnessAlias &l, const WitnessAlias &r) {
    return l.partialColoring == r.partialColoring;
}

friend bool is_less_implementation(const WitnessAlias &l, const WitnessAlias &r) {
    return l.partialColoring < r.partialColoring;
}
\end{code}

\paragraph{Relabeling.}
The \texttt{relabel\_implementation} function takes as input an injective mapping whose domain is the set of active labels occurring in the local witness that is currently being processed. It returns a new local witness where each active label occurring in it is replaced by its image under the mapping. This function is relevant when using the symmetry-breaking search strategy, as it allows the algorithm to identify local witnesses that differ only by a change of labels \cite{de2024state}.

In our example, this function iterates through the color classes (each represented as a set of labels) and applies the relabeling accordingly. The resulting witness will have each original label replaced with its corresponding value from the mapping.

\begin{code}[language=C++]
WitnessAlias relabel_implementation(const std::map<unsigned, unsigned> &relabelingMap) const {
    auto relabeled = WitnessAlias();
    for (const auto &cell : partialColoring) {
        std::set<unsigned> relabeledCell;
        for (const auto &vertex : cell) {
            relabeledCell.insert(relabelingMap.at(vertex));
        }
        relabeled.partialColoring.push_back(relabeledCell);
    }
    std::sort(relabeled.partialColoring.begin(), relabeled.partialColoring.end());
    return relabeled;
}
\end{code}

\paragraph{Hash.} The \texttt{hash} function is designed to generate a unique hash value for a given local witness.
This mechanism is crucial in dynamic programming algorithms, where rapid identification of equivalent local witnesses is essential.
In our example, it iterates through each vertex label in every color class and combines their values to produce the final hash. To distinguish between different color classes, the implementation inserts the unsigned sentinel value \verb|-1u| as a delimiter during the hashing process.

This ensures that distinct local witnesses, which might otherwise produce the same hash value, can be distinguished. For example, without this delimiter, the hash function could generate the same hash value for $\{\{1\}, \{2,3\}\}$ and $\{\{1,2\}, \{3\}\}$.

\begin{code}[language=C++]
void hash(Hasher &h) const override {
    for (const auto &s : partialColoring) {
        for (const auto &c : s)
            h << c;
        h << -1u;
    }
}
\end{code}

\paragraph{Printable Information.} For debugging and presentation purposes, the witness structure has a method called \verb|witness_info| that is used to print the attributes of a witness in a human-readable way.
In our example, this method traverses each color class in the \texttt{partialColoring} attribute and prints the vertex labels contained within the color class.

\begin{code}[language=C++]
void witness_info(std::ostream &os) const {
    os << "partialColoring:";
    for (const auto &cell : partialColoring) {
        os << "{ ";
        for (const auto &item : cell) {
            os << item;
            if (item != *cell.crbegin())
                os << ",";
        }
        os << "}";
    }
    os << "\n";
}
\end{code}

\subsection{Defining the Transition Functions of a DP-Core}
This section demonstrates how to implement the transition functions of a DP-core in $\treewidzard$. The following code outlines the methods that need to be implemented for a DP-core.

\begin{code}[language=C++, label={code:dpcore}, caption={Methods in a DP-Core}]
struct ExampleDynamicCore : CoreWrapper<ExampleDynamicCore, ExampleWitness, WitnessSetTypeTwo> {
    static std::map<std::string, std::string> metadata() {
        return {
            {"CoreName", "name"},
            {"CoreType", "core_type"},
            {"ParameterType", "parameter_type"},
        };
    }

    ExampleDynamicCore(const parameterType &parameters);

    void initialize_leaf(WitnessSet &witnessSet);

    void intro_v_implementation(unsigned int i, const Bag &, const WitnessAlias &w, WitnessSet &witnessSet);

    void intro_e_implementation(unsigned int i, unsigned int j, const Bag &, const WitnessAlias &w, WitnessSet &witnessSet);

    void forget_v_implementation(unsigned int i, const Bag &, const WitnessAlias &w, WitnessSet &witnessSet);

    void join_implementation(const Bag &, const WitnessAlias &w1, const WitnessAlias &w2, WitnessSet &witnessSet);

    bool is_final_witness_implementation(const Bag &, const WitnessAlias &w);

    void clean_implementation(WitnessSet &);

    int inv_implementation(const Bag &, const WitnessAlias &w);
};
\end{code}

Next, we illustrate the implementation of the methods listed above by defining a DP-core that decides whether a given graph has chromatic number at most $c$, for a specified value of $c$.

The color bound $c$, i.e., the number of colors available to color the vertices of the graph, is stored in the implementation as the unsigned field \verb|k|.
The method \verb|metadata()| encapsulates metadata for the DP-core. The core is named \verb|ChromaticNumber_AtMost| and has a type of \texttt{Bool}, meaning that the output produced by the core is either $1$ or $0$, depending on whether the graph is $c$-colorable. The metadata declares an unsigned integer parameter; the constructor unpacks the parser value and stores it in this field.

\begin{code}[language=C++]
struct ChromaticNumber_AtMost_Core : CoreWrapper<ChromaticNumber_AtMost_Core, ChromaticNumber_AtMost_Witness, WitnessSetTypeTwo>
{
    unsigned k;

    static std::map<std::string, std::string> metadata() {
        return {
            {"CoreName", "ChromaticNumber_AtMost"},
            {"CoreType", "Bool"},
            {"ParameterType", "UnsignedInt"},
        };
    }

    ChromaticNumber_AtMost_Core(const parameterType &parameters) {
        auto [n] = unpack_typed_args<int>(parameters);
        k = static_cast<unsigned>(n);
    }
};
\end{code}

\subsubsection{Auxiliary Functions}

Next, we describe three auxiliary functions that are used to facilitate the implementation of the transition functions of a DP-core.
The first auxiliary function is used to clone a local witness. Cloning is particularly
useful when we want to modify a copy of a local witness without altering the original one.

\begin{code}[language=C++]
auto witness_prime = witness.clone();
\end{code}

The next command is used to create a new local witness with default attributes.
\begin{code}[language=C++]
auto witness = make_shared<WitnessAlias>();
\end{code}

Lastly, we often need to insert a witness into a witness set. This can be done using the following command:

\begin{code}[language=C++]
witnessSet.insert(witness);
\end{code}

\subsubsection{Initialization at Leaf Nodes}
Dynamic programming cores process a tree decomposition from the leaves towards the root. More specifically, the algorithm starts by assigning a witness set to each leaf node. In $\treewidzard$, this is done by the \texttt{initialize\_leaf} function.

The graph corresponding to each leaf node is the empty graph, that is, the graph with no vertices. The only valid coloring of this graph is the empty coloring. For this reason, in the case of the DP-core \verb|ChromaticNumber_AtMost|, the initial witness set contains exactly one local witness, whose \verb|partialColoring| attribute is empty.

\begin{code}[language=C++]
void initialize_leaf(WitnessSet &witnessSet) {
    auto initial_witness = make_shared<WitnessAlias>();
    witnessSet.insert(initial_witness);
}
\end{code}

\subsubsection{Introducing a New Vertex}

In $\treewidzard$, the \verb|intro_v_implementation| function handles the case where a new vertex with label $i$ is introduced at a given node of a tree decomposition.
This function receives as input a label \codeinline{i}, a reference \codeinline{b} to a set of active labels, a reference \codeinline{w}
to a witness, and a reference \codeinline{witnessSet} to an empty witness set. We note that the label \codeinline{i} is assumed not to be present in
\codeinline{b}.

In our particular example, the introduction of a new vertex requires extending the current partial coloring by considering all valid ways of assigning a color for the newly introduced vertex.
More specifically, the algorithm iterates over each color class in the \verb|partialColoring| attribute, clones the witness, adds the new vertex to the color class, and then inserts the updated witness into the witness set.
Additionally, if the number of color classes is less than the stored color bound, a new witness is also created in which the new vertex forms its own color class. This guard is the key local invariant ensuring that every generated witness represents at most $c$ color classes.

\begin{code}[language=C++]
void intro_v_implementation(unsigned i, const Bag &, const WitnessAlias &w, WitnessSet &witnessSet) {
    for (size_t idx = 0; idx < w.partialColoring.size(); ++idx) {
        auto witness = w.clone();
        witness->partialColoring[idx].insert(i);
        std::sort(witness->partialColoring.begin(), witness->partialColoring.end());
        witnessSet.insert(std::move(witness));
    }
    if (w.partialColoring.size() < this->k) {
        auto witness = w.clone();
        std::set<unsigned> iCell = {i};
        witness->partialColoring.push_back(iCell);
        std::sort(witness->partialColoring.begin(), witness->partialColoring.end());
        witnessSet.insert(std::move(witness));
    }
}
\end{code}

\subsubsection{Introducing a New Edge}

The function \verb|intro_e_implementation| specifies the action of the DP-core on a local witness when processing an instruction that introduces an edge between vertices labeled \codeinline{i} and \codeinline{j}. This function receives as input labels \codeinline{i} and \codeinline{j}, a reference \codeinline{b} to a set of active labels, a reference \codeinline{w} to a witness, and a reference \codeinline{witnessSet} to an empty witness set. We note that it is assumed that both \codeinline{i} and \codeinline{j} are present in \codeinline{b}.

In our example, we need to ensure that the new edge's endpoints are placed in different color classes, as this is a key constraint of graph coloring. If both endpoints of the edge belong to the same color class, the function terminates early and discards the current witness, as it violates the coloring constraint. If the endpoints are in different color classes, the witness is cloned and added to the witness set.

\begin{code}[language=C++]
void intro_e_implementation(unsigned int i, unsigned int j, const Bag &, const WitnessAlias &w, WitnessSet &witnessSet) {
    for (const auto &cell : w.partialColoring) {
        if (cell.count(i) && cell.count(j))
            return;
    }
    witnessSet.insert(w.clone());
}
\end{code}

\subsubsection{Forgetting a Vertex}
The function \texttt{forget\_v\_implementation} is called when a vertex is forgotten.
This function receives as input a label \codeinline{i}, a reference \codeinline{b} to a set of active labels, a reference \codeinline{w}
to a witness, and a reference \codeinline{witnessSet} to an empty witness set. We note that the label \codeinline{i} is assumed to be present in \codeinline{b}.

In our example, the function locates the color class containing the label, clones the witness, and then removes the label from the identified class. If the color class becomes empty after the removal, the implementation removes that cell from the vector and sorts the remaining cells to keep the canonical representation.

\begin{code}[language=C++]
void forget_v_implementation(unsigned int i, const Bag &, const WitnessAlias &w, WitnessSet &witnessSet) {
    for (size_t idx = 0; idx < w.partialColoring.size(); ++idx) {
        if (w.partialColoring[idx].count(i)) {
            auto witness = w.clone();
            witness->partialColoring[idx].erase(i);
            if (witness->partialColoring[idx].empty()) {
                witness->partialColoring.erase(
                    witness->partialColoring.begin() + static_cast<ptrdiff_t>(idx));
            }
            std::sort(witness->partialColoring.begin(), witness->partialColoring.end());
            witnessSet.insert(std::move(witness));
            break;
        }
    }
}
\end{code}

\subsubsection{Join Operation}
The \verb|join_implementation| function is called when the algorithm processes a join node.
This function receives a reference to a set of active labels \codeinline{b}, references to two witnesses \codeinline{w1} and \codeinline{w2}, and a reference to an empty local witness set \codeinline{witnessSet}. In this function, new witnesses are generated by joining \codeinline{w1} and \codeinline{w2}, and inserted into \codeinline{witnessSet}.

In our example, the \texttt{join\_implementation} function checks if the partial colorings of the two witnesses are identical, which indicates that the partial solutions are consistent.
If this is the case, a clone of the first witness is inserted into the new witness set.

\begin{code}[language=C++]
void join_implementation(const Bag &, const WitnessAlias &w1, const WitnessAlias &w2, WitnessSet &witnessSet) {
    if (w1.partialColoring == w2.partialColoring) {
        witnessSet.insert(w1.clone());
    }
}
\end{code}

\subsubsection{Final Witness Verification}
The \texttt{is\_final\_witness\_implementation} function takes as input a reference \codeinline{b} to a set of active labels and a reference \codeinline{w}
to a witness and returns true if and only if \verb|w| is a final witness.

Each local witness is a final witness in the DP-core \verb|ChromaticNumber_AtMost|. Therefore, in this case, the function always returns true.
\begin{code}[language=C++]
bool is_final_witness_implementation(const Bag &, const WitnessAlias &) {
    return true;
}
\end{code}

\subsubsection{Clean Function}
In some DP-cores, it is possible to identify witnesses that are redundant while processing a tree decomposition. Such witnesses can be removed without changing whether a solution will be found. In $\treewidzard$, this cleaning process is performed using the \verb|clean_implementation| function. For \verb|ChromaticNumber_AtMost|, this function is implemented as the identity function and does not modify the input \verb|witnessSet|.
\begin{code}[language=C++]
void clean_implementation(WitnessSet &) {
}
\end{code}

\section{Example DP-Core: \texttt{IndependenceNumber}}
\label{section:independent set}

We now turn to a second example DP-core, parallel in structure to the previous one but illustrating the optimization variant of the framework. Here we introduce the \verb|IndependenceNumber| DP-core, which computes the size of the largest independent set in graphs of bounded treewidth. Its implementation uses the C++ structures \verb|IndependentSet_Max_Witness| and \verb|IndependentSet_Max_Core|. Unlike the \verb|ChromaticNumber_AtMost| problem discussed earlier, which is a decision problem, this is an optimization problem aiming to find the best numeric solution.

A key distinction between \verb|ChromaticNumber_AtMost| and \verb|IndependenceNumber| lies in the type of DP-core: the former is a decision core, while the latter is an optimization core. As such, \verb|IndependenceNumber| requires a nontrivial invariant function to guide the optimization process. Additionally, it does not take a fixed target value as input; the optimal value depends on the structure of the input graph.

The optimization is managed by the \verb|inv_implementation| function, which assigns a numeric value to each local witness based on its contents. At the root node of the decomposition, $\treewidzard$ evaluates all local witnesses and selects the optimal one, either the minimum or maximum, depending on whether the DP-core is defined as \verb|Min| or \verb|Max|. For \verb|IndependenceNumber|, the final result is the maximum value assigned to any root witness, representing the size of the maximum independent set.

While all DP-cores include an \verb|inv_implementation| function, in decision cores it typically reports the trivial $0$/$1$ value induced by acceptance; witness validity itself is determined by \verb|is_final_witness_implementation|. In optimization cores like \verb|IndependenceNumber|, \verb|inv_implementation| assigns meaningful numeric values used to select an optimum.

Despite being an optimization DP-core, the concept of a final local witness remains relevant. In $\treewidzard$, the \verb|is_final_witness_implementation| function determines whether a witness at the root should be considered valid for final evaluation.

\subsection{Local Witness}

We define a local witness using a structure named \verb|IndependentSet_Max_Witness|, which encapsulates the necessary attributes and required functions such as comparison, relabeling, hashing, and printing. The design is based on the idea that a local witness stores information that contributes to constructing an independent set, associated with a node in an instructive tree decomposition (ITD).

\paragraph{Attributes.} Each local witness maintains a set \verb|used| of active labels that have been included in the independent set. Additionally, the variable \verb|size| tracks the total number of vertices that have been included so far, accounting for both the currently active labels and those that have been forgotten during the dynamic programming process.
Local witnesses store only vertices currently relevant (active labels), omitting explicitly forgotten vertices, as their contribution is already counted and retained numerically in the \verb|size| attribute. Since forgotten labels are no longer present at the current node, the local witness does not explicitly represent a full independent set, but rather retains sufficient information to continue constructing one as the computation proceeds.

\begin{code}[language=C++]
struct IndependentSet_Max_Witness : WitnessWrapper<IndependentSet_Max_Witness>
{
    int size = 0;
    std::set<unsigned> used;
};
\end{code}

\paragraph{Comparison Functions.}
The \verb|is_equal_implementation| function checks whether two witnesses have equal \verb|size| and \verb|used| values. The \verb|is_less_implementation| function defines a strict ordering: it first compares the \verb|size| attribute, and if those are equal, it compares the \verb|used| attribute.

\begin{code}[language=C++]
friend bool is_equal_implementation(const WitnessAlias &l, const WitnessAlias &r) {
    return l.size == r.size && l.used == r.used;
}

friend bool is_less_implementation(const WitnessAlias &l, const WitnessAlias &r) {
    if (l.size != r.size) return l.size < r.size;
    if (l.used != r.used) return l.used < r.used;
    return false;
}
\end{code}

\paragraph{Relabeling.}
The \verb|relabel_implementation| method returns a witness whose \verb|size| attribute is the same as that of the current witness, and whose \verb|used| attribute is the image of the current witness's \verb|used| attribute under the map \verb|relabelingMap|.

\begin{code}[language=C++]
WitnessAlias relabel_implementation(const std::map<unsigned, unsigned> &relabelingMap) const {
    auto relabeled = WitnessAlias();
    relabeled.size = size;
    for (unsigned i : used) relabeled.used.insert(relabelingMap.at(i));
    return relabeled;
}
\end{code}

\paragraph{Hash.}
The hash function combines the \verb|size| value with the elements of the \verb|used| set to identify each local witness uniquely. Because \verb|std::set| stores elements in sorted order, structurally identical sets produce consistent hashes. The inclusion of \verb|size| ensures that local witnesses with the same set but different sizes are distinguished.
\begin{code}[language=C++]
void hash(Hasher &h) const override {
    h << size;
    for (unsigned i : used) h << i;
}
\end{code}

\paragraph{Printable Information.}
The \verb|witness_info| function displays the size of the included vertices in the local witness and lists the labels currently present in the \verb|used| set, providing a clear description of the local witness in a human-readable format.

\begin{code}[language=C++]
void witness_info(std::ostream &os) const {
    os << "Independent set of size " << size << " using";
    for (unsigned i : used) os << ' ' << i;
    os << '\n';
}
\end{code}

\subsection{Defining the Transition Functions of a DP-Core}

The transition functions in \verb|IndependenceNumber| follow the same structure and behavior as described in the previous section for decision DP-cores. They update the local witness attributes \verb|used| and \verb|size| based on the semantics of the independent set problem and the structure of the tree decomposition.

The key adaptation in this optimization DP-core is the use of a nontrivial \verb|inv_implementation| method, which returns the \verb|size| attribute of each local witness. This numeric invariant is used by $\treewidzard$ to compare witnesses and select the optimal value (in this case, the maximum size) at the root.

No further modifications to the transition logic are needed. As long as the witness structure and its attributes are correctly updated by the transition functions, $\treewidzard$ will propagate and compare witnesses based on the numeric invariant.

To begin the core definition, we specify the required metadata in the \verb|metadata()| method. This includes the core's name, its type (\verb|Max| to indicate that the invariant should be maximized), and the parameter type (\verb|None|), since this problem does not require any input parameters. The constructor enforces this by asserting that the parameter list is empty.

\begin{code}[language=C++]
struct IndependentSet_Max_Core: CoreWrapper<IndependentSet_Max_Core, IndependentSet_Max_Witness, WitnessSetTypeTwo> {
    static std::map<std::string, std::string> metadata() {
        return {
            {"CoreName", "IndependenceNumber"},
            {"CoreType", "Max"},
            {"ParameterType", "None"},
        };
    }

    IndependentSet_Max_Core(const parameterType &parameters) {
        assert(parameters.size() == 0);
        (void)parameters;
    }
};
\end{code}

\subsubsection{Initialization at Leaf Nodes}

In $\treewidzard$, the process begins at the leaf nodes. The graph associated with a leaf node is the empty graph, which trivially admits the empty independent set. Consequently, the \verb|initialize_leaf| function creates a witness set with a single local witness, where the attribute \verb|size| is zero and the attribute \verb|used| is empty.
In the code, the expression \verb|std::make_shared<WitnessAlias>()| constructs this local witness, which is then inserted into the witness set passed as a parameter to the function.

\begin{code}[language=C++]
void initialize_leaf(WitnessSet &witnessSet) {
    witnessSet.insert(std::make_shared<WitnessAlias>());
}
\end{code}

\subsubsection{Introducing a New Vertex}
When a new vertex is introduced, two possibilities are considered. One possibility is to leave the vertex out of the independent set, in which case the current independent set remains unchanged. The other possibility is to include the vertex in the independent set. At the time of introduction, the vertex is isolated because no edges involving it have been introduced yet. This ensures that including the vertex does not violate the independence condition.

The function \verb|intro_v_implementation| handles these two cases by generating two local witnesses based on the input witness. A copy of the input witness is added to the witness set to represent the case where the label of the introduced vertex is ignored. A second copy is updated by adding the introduced label to the \verb|used| set and incrementing the \verb|size| by one. This updated local witness reflects the case where the current local witness includes the label of the introduced vertex.

\begin{code}[language=C++]
void intro_v_implementation(unsigned i, const Bag &, const WitnessAlias &w, WitnessSet &witnessSet) {
    witnessSet.insert(w.clone());

    auto wPrime = w.clone();
    ++wPrime->size;
    wPrime->used.insert(i);
    witnessSet.insert(wPrime);
}
\end{code}

\subsubsection{Introducing a New Edge}
When an edge is introduced between two vertices, the current independent set must be checked to ensure that it still satisfies the independence condition. Specifically, if both endpoints of the edge are already included in the independent set, the configuration violates the independence requirement and must be discarded.

To enforce this constraint, the function \verb|intro_e_implementation| checks whether a given local witness remains valid after the edge is introduced. The parameters \verb|i| and \verb|j| represent the labels of the two endpoints of the new edge. If both \verb|i| and \verb|j| are present in the \verb|used| set, the local witness is considered invalid and is excluded simply by not inserting it into the witness set. Otherwise, a copy of the input local witness is added, indicating that it remains valid after the edge is introduced.

\begin{code}[language=C++]
void intro_e_implementation(unsigned i, unsigned j, const Bag &, const WitnessAlias &w, WitnessSet &witnessSet) {
    if (w.used.count(i) && w.used.count(j))
        return;
    witnessSet.insert(w.clone());
}
\end{code}

\subsubsection{Forgetting a Vertex}

In our example, forgetting a label corresponds to removing it from the set of active labels. Although the label is no longer active, its earlier inclusion in the independent set remains recorded in the local witness. This enables the local witness to preserve essential information while adapting to a smaller set of active labels.

The function \verb|forget_v_implementation| performs this update. It clones the input local witness, removes the label \verb|i| from the \verb|used| set, and inserts the modified witness into the witness set. The \verb|size| attribute remains unchanged, since forgetting a label does not alter the fact that the corresponding vertex was part of the independent set earlier in the process.

\begin{code}[language=C++]
void forget_v_implementation(unsigned i, const Bag &, const WitnessAlias &w, WitnessSet &witnessSet) {
    auto wPrime = w.clone();
    wPrime->used.erase(i);
    witnessSet.insert(wPrime);
}
\end{code}

\subsubsection{Join Operation}

At a join node, two partial independent-set witnesses are compatible only if they agree on the active bag labels selected into the independent set. In our setting, each local witness represents a partial independent set and maintains two key components: the size of the independent set so far (\verb|size|), and the set of currently active labels included in the set (\verb|used|).

Forgotten vertices (those no longer active in the bag) can still be part of the independent set, but they are processed in separate branches of the tree decomposition before the join. Since no edge can exist between forgotten vertices from different branches, we only need to check for conflicts among the active labels during the join.

A conflict arises if the union of the two \verb|used| sets includes both endpoints of an edge introduced in either branch. To prevent such violations, the DP-core allows a join only when the two input witnesses have identical \verb|used| sets. If they differ, a conflict may arise, and the function returns without inserting anything into the witness set.

If the \verb|used| sets are equal, the join is guaranteed to be safe. The resulting witness represents the combined independent set, with its size computed as the sum of the sizes of the two input witnesses minus the number of shared active labels (to avoid double-counting). A clone of \verb|w1| is then created, its \verb|size| attribute updated, and the resulting witness inserted into the witness set.

\begin{code}[language=C++]
void join_implementation(const Bag &, const WitnessAlias &w1, const WitnessAlias &w2, WitnessSet &witnessSet) {
    if (w1.used != w2.used)
        return;
    int joinedCount = w1.size + w2.size - static_cast<int>(w1.used.size());
    auto wPrime = w1.clone();
    wPrime->size = joinedCount;
    witnessSet.insert(wPrime);
}
\end{code}

\subsubsection{Final Witness Verification}
To determine whether a local witness is valid at the root of the decomposition, the DP-core uses the function \verb|is_final_witness_implementation|. Since each local witness is constructed to represent a valid independent set, all witnesses are considered acceptable at the root. Therefore, the function simply returns \verb|true| for any given local witness.

\begin{code}[language=C++]
bool is_final_witness_implementation(const Bag &, const WitnessAlias &) {
    return true;
}
\end{code}

\subsubsection{Clean Function}
The DP-core \verb|IndependenceNumber| is designed to compute the maximum size of an independent set in a graph. Many local witnesses are generated during the dynamic programming process, but some may be redundant. For instance, consider two local witnesses with the same \verb|used| set but different \verb|size| values. Since the \verb|used| sets are identical, any further extension from these local witnesses will proceed identically and produce local witnesses with the same \verb|used| sets. In this case, retaining only the local witness with the larger \verb|size| is more efficient.

In $\treewidzard$, this redundancy is resolved by the \verb|clean_implementation| function. If no such function is implemented, a default identity function is used that leaves the \verb|witnessSet| unchanged. In this implementation, when multiple local witnesses share the same \verb|used| set, only the one with the largest \verb|size| is kept. The \verb|clean_implementation| function removes the rest, improving efficiency without affecting correctness.

\begin{code}[language=C++]
void clean_implementation(WitnessSet &witnessSet) {
    auto witnesses = std::vector<std::shared_ptr<WitnessAlias>>();
    for (const auto &w : witnessSet)
        witnesses.push_back(std::dynamic_pointer_cast<WitnessAlias>(w));

    std::sort(witnesses.begin(), witnesses.end(),
              [](const auto &w0, const auto &w1) {
                  if (w0->used != w1->used) {
                      return w0->used < w1->used;
                  }
                  return w0->size > w1->size;
              });

    witnesses.erase(std::unique(witnesses.begin(), witnesses.end(),
                                [](const auto &w0, const auto &w1) {
                                    return w0->used == w1->used;
                                }),
                    witnesses.end());

    witnessSet = WitnessSet();
    for (auto &w : witnesses)
        witnessSet.insert(std::move(w));
}
\end{code}

Note that the input to \verb|clean_implementation| is a generic witness set. Therefore, one should use explicit casting (such as \verb|std::dynamic_pointer_cast|) to convert witnesses to the specific witness type before accessing their defined attributes.

\subsubsection{The Invariant Function}

A graph invariant is a property that remains unchanged under graph isomorphisms and is typically expressed as a single value. In $\treewidzard$, computing a graph invariant requires implementing the function \verb|inv_implementation|. This function takes a local witness as input and returns an integer value representing its contribution to the invariant. Intuitively, \verb|inv_implementation| assigns a value to each local witness, and $\treewidzard$ evaluates all such values at the root node. Based on the declared DP-core type—either \verb|Max| or \verb|Min|—it selects the corresponding extremal value as the graph's invariant.

In the case of \verb|IndependenceNumber|, the core type is \verb|Max|, so $\treewidzard$ selects the maximum value among all local witnesses at the root. The \verb|inv_implementation| function returns the \verb|size| attribute of the local witness, directly reflecting the size of the independent set represented by the witness.

\begin{code}[language=C++]
int inv_implementation(const Bag &, const WitnessAlias &w) {
    return w.size;
}
\end{code}

\section{Conclusion}
\treewidzard turns decomposition-based dynamic programs into reusable, compositional components. The same components are used both for instance-level evaluation and for bounded-width automated
reasoning. The central abstraction is the notion of a DP-core: a finite witness representation together with one local
transition rule for each ITD constructor, plus a final-witness test and optional dominance pruning and
invariant extraction. 

This viewpoint connects parameterized algorithm design with automated reasoning over bounded-width graph classes. In particular, the bounded-width validity procedure ranges over all $k$-instructive tree decompositions and
searches the induced global state space, returning either a concrete counterexample graph or (when the
search closes) a successful verdict that the specification holds for all graphs of width at most $k$ under the DP-core coherence assumptions inherited from the width-based ATP framework~\cite{de2023width,de2024state}. Concretely, these are the standard requirements that each DP-core be isomorphism-invariant (acceptance and invariant values depend only on the underlying graph, not on the chosen ITD representation) and that its witness sets remain finite on width-$\le k$ inputs; under these assumptions the bounded-width procedure is sound and refutationally complete. State
canonization and premise-based pruning reduce the explored state space on typical ATP specifications and
implications while preserving these guarantees in the bounded-width setting.

We expect \treewidzard to be useful as a laboratory for the development of width-based dynamic programming algorithms and for the investigation
of graph-theoretic conjectures on classes of graphs of bounded treewidth/pathwidth. More specifically, DP-cores let one package familiar width-based algorithms as composable modules, and then reuse these modules across concrete instances and across bounded-width validity questions.

\bibliographystyle{plain}  
\bibliography{references}

\appendix
\section{Input File Formats}\label{appendix:file_formats}

This appendix provides a comprehensive specification of the file formats required for $\treewidzard$ operation. The system supports two distinct modes of graph specification corresponding to different usage scenarios. For model checking applications where the graph is known explicitly, users provide either a graph file paired with its tree decomposition or a single instructive tree decomposition encoding both structures simultaneously. For automated theorem proving applications that explore all graphs within treewidth bounds, only the property specification is required. All operational modes require a property file defining the graph-theoretic conditions to be verified, whose format we describe in Appendix~\ref{appendix:FormatPropertyFile}.

\subsection{Graph File Format}\label{appendix:graph_file}

The graph input format extends the DIMACS format standardized by the PACE Challenge 2017~\cite{dell2018pace}, providing a text-based representation suitable for simple graphs and multigraphs. A graph file carries the \verb|.gr| extension and consists of comment lines, a header line specifying graph parameters, and edge specification lines.

Comment lines begin with the character \verb|c| and may appear at any position within the file. The system ignores all content on comment lines, enabling documentation and annotation without affecting parsing. The header line, which must be the first non-comment line in the file, begins with the character sequence \verb|p tw| followed by two integers: the number of vertices $n$ and the number of edges $m$, separated by whitespace. Vertices are labeled consecutively from $1$ to $n$, establishing a canonical labeling that all subsequent edge specifications must respect.

Each edge line specifies the \emph{edge relation} on the vertex set $V=\{1,\dots,n\}$.
Concretely, a line containing two integers $u\ v$ asserts that the undirected edge relation holds
between $u$ and $v$ (equivalently, $E(u,v)$ and $E(v,u)$). The format permits isolated vertices and,
syntactically, repeated edge lines. Repeated lines are interpreted as parallel edges when \treewidzard
prints counterexamples (where edges may be shown with explicit identifiers), and they may be ignored
if one restricts attention to simple-graph semantics. Self-loops ($u=v$) are not supported.

The following example illustrates the format through a path graph $P_5$ containing five vertices and four edges:

\begin{code}
c This file describes a path with five vertices and four edges.
p tw 5 4
1 2
2 3
c we are half-way done with the instance definition.
3 4
4 5
\end{code}

The header line \verb|p tw 5 4| declares a graph with five vertices labeled $1, 2, 3, 4, 5$ and four edges. The four subsequent edge specifications construct the path $1$--$2$--$3$--$4$--$5$. Note that comment lines may be interspersed with edge specifications, providing documentation at any point in the file structure.

\subsection{Tree Decomposition File Format}\label{appendix:decomposition_file}

Tree decompositions are specified through files with the \verb|.td| extension, following the format established by the PACE Challenge 2017~\cite{dell2018pace}. A tree decomposition $(T, \mathcal{B})$ of a graph $G$ consists of a tree $T$ whose nodes are indexed by integers and a collection $\mathcal{B} = \{B_i\}_{i \in V(T)}$ of bags, where each bag $B_i \subseteq V(G)$ contains a subset of the graph's vertices. The decomposition must satisfy three conditions: every edge of $G$ appears in at least one bag, every vertex of $G$ appears in at least one bag, and for each vertex $v \in V(G)$, the set of bags containing $v$ induces a connected subtree of $T$. The width of the decomposition is defined as $\max_{i \in V(T)} |B_i| - 1$, one less than the size of the largest bag.

The file format parallels the graph format structure: comment lines begin with \verb|c|, a header line specifies decomposition parameters, bag content lines define the bag collection, and tree edge lines specify the tree structure. The header line begins with \verb|s td| followed by three integers: the number of bags $N$, the maximum bag size $w$, and the number of vertices $n$ in the original graph. The width of this decomposition equals $w - 1$. Each bag is specified through a line beginning with \verb|b|, followed by the bag index (an integer in $[1, N]$) and the space-separated list of vertices contained in that bag. Vertices are referenced using the same labels established in the corresponding graph file. Empty bags are permitted and represented by a bag specification line containing only the bag index. Each bag index must appear exactly once in a bag specification line. The tree structure is defined through edge specifications: each tree edge is represented by a line containing two bag indices, establishing that those bags are adjacent in the tree $T$. The collection of tree edges must form a connected acyclic graph (i.e., a tree) on the $N$ bags.

The following example specifies a tree decomposition of width 2 for the path graph $P_5$:

\begin{code}
c This file describes a tree decomposition with 4 bags, width 2, for a graph with 5 vertices
s td 4 3 5
b 1 1 2 3
b 2 2 3 4
b 3 3 4 5
b 4
1 2
2 3
2 4
\end{code}

The header line \verb|s td 4 3 5| declares a decomposition with four bags, maximum bag size 3 (implying width 2), for a graph with five vertices. The four bags are: $B_1 = \{1, 2, 3\}$, $B_2 = \{2, 3, 4\}$, $B_3 = \{3, 4, 5\}$, and $B_4 = \emptyset$ (the empty bag). The tree structure is defined through three edges: $(1, 2)$, $(2, 3)$, and $(2, 4)$, forming a tree where bag 2 serves as the root with three children. This decomposition satisfies the tree decomposition properties: every edge of the path appears in at least one bag, every vertex appears in consecutive bags along a path in the tree, and the maximum bag size is 3.

\subsection{Instructive Tree Decomposition Format}\label{appendix:itd_file}

Instructive tree decompositions (ITDs) provide an alternative encoding that specifies both a graph and its tree decomposition through a sequence of construction operations. Rather than listing bags and edges explicitly, an ITD file describes how to build the decomposition incrementally through a tree of instructions. Each instruction node in this tree performs one of several operations: introducing a new vertex label, introducing an edge between existing labels, forgetting a vertex label, or joining two sub-decompositions.

An ITD file consists of lines, each specifying a single instruction node. Every line begins with a unique node identifier (a positive integer) followed by the operation name and, for non-leaf nodes, references to child nodes. Leaf nodes, which represent the base case of empty decompositions, are specified simply as:
\begin{center}
\verb|<Node ID>  Leaf|
\end{center}
Unary operations that modify a single child decomposition follow the format:
\begin{center}
\verb|<Node ID>  <Operation>(<Child Node ID>)|
\end{center}
where the operation is one of \verb|IntroVertex_i| (introducing vertex label $i$), \verb|IntroEdge_i_j| (introducing an edge between labels $i$ and $j$), or \verb|ForgetVertex_i| (removing label $i$ from the active label set). Binary join operations that combine two sub-decompositions use the format:
\begin{center}
\verb|<Node ID>  Join(<Child1 Node ID>, <Child2 Node ID>)|
\end{center}
The following example illustrates a $2$-ITD encoding the \emph{diamond graph} (two triangles sharing an edge):

\begin{instructivedecompositionfile}
1 Leaf
2 IntroVertex_1(1)
3 IntroVertex_2(2)
4 IntroEdge_1_2(3)
5 IntroVertex_3(4)
6 IntroEdge_2_3(5)
7 Leaf
8 IntroVertex_1(7)
9 IntroVertex_2(8)
10 IntroVertex_3(9)
11 IntroEdge_1_3(10)
12 Join(6,11)
13 ForgetVertex_2(12)
14 IntroVertex_2(13)
15 IntroEdge_2_3(14)
16 IntroEdge_1_2(15)    
\end{instructivedecompositionfile}

The ITD construction proceeds through two branches that are subsequently joined. The first branch begins at the leaf node 1 and constructs a path $1$--$2$--$3$ through operations 2--6: node 2 introduces label 1, node 3 introduces label 2, node 4 adds the edge $(1,2)$, node 5 introduces label 3, and node 6 adds the edge $(2,3)$. The second branch similarly constructs a different subgraph through operations 7--11, introducing labels 1, 2, and 3 and adding the edge $(1,3)$. The join operation at node 12 combines these branches, requiring that both have identical active label sets (which they do: both have labels $\{1, 2, 3\}$ active after their respective construction sequences). The subsequent operations 13--16 complete the construction by manipulating label 2 and adding the final edges. This instruction sequence implicitly defines both a graph and a valid tree decomposition of width 2, demonstrating the ITD's ability to encode structural information through operations rather than explicit listings.

\subsection{Property File Format}
\label{appendix:FormatPropertyFile}

Property files define the graph-theoretic conditions that $\treewidzard$ verifies through DP-core evaluation. Every computational task performed by $\treewidzard$—whether model checking a specific graph, exploring the space of bounded treewidth graphs for automated theorem proving, or computing graph invariants—requires a property file specifying the verification criterion. A property file consists of three components: variable declarations binding identifiers to DP-cores, an algebraic formula expressing the property to be verified, and optional comments documenting the specification.

\subsubsection{Variable Declaration and DP-Core Binding}

Variable declarations establish the connection between symbolic identifiers and DP-core implementations. Each declaration follows the syntax:
\begin{center}
\verb|<var> := <CoreName>(<parameter>)|
\end{center}
where \verb|<var>| is an alphanumeric identifier, \verb|<CoreName>| specifies the DP-core implementation to instantiate, and \verb|<parameter>| provides the core's configuration parameter. For DP-cores that require no parameterization, the parentheses remain empty: \verb|<var> := <CoreName>()|. The semantics of variable binding are as follows: during tree decomposition traversal, the specified DP-core processes the decomposition bottom-up, maintaining witness sets at each node. Upon reaching the root, the variable receives a Boolean value indicating acceptance (value 1) or rejection (value 0) based on whether any final witness exists in the root's witness set.

Every variable is Boolean: a variable $x$ records whether the graph was accepted by the DP-core assigned to $x$. The expression \verb|INV(x)| accesses the numerical value reported by that core. For decision DP-cores that merely determine property membership, this value equals the Boolean value: \verb|INV(x)| = 1 if the core accepts and 0 otherwise. DP-cores computing invariants override this default through custom invariant implementations, enabling computation of graph parameters such as independence number or vertex count. The distinction between the Boolean acceptance value and the numerical invariant value provides flexibility: a DP-core can accept all graphs (always returning 1) while computing meaningful invariants whose values depend on graph structure.

Consider the following variable declarations:

\begin{code}
x := ChromaticNumber_AtMost(3)
y := IndependenceNumber()
\end{code}

The first declaration binds \verb|x| to a decision DP-core that accepts graphs that are $3$-colorable. For this core, the invariant \verb|INV(x)| equals the Boolean value: 1 for $3$-colorable graphs and 0 otherwise. The second declaration binds \verb|y| to an invariant-computing DP-core that accepts all graphs but computes the independence number as its invariant. Thus \verb|y| always equals 1 (all graphs are accepted), while \verb|INV(y)| provides the size of the maximum independent set.

\subsubsection{Formula Specification and Evaluation Semantics}

Following the variable declarations, the property file must contain a formula section beginning with the keyword \verb|Formula| on a line by itself. All subsequent non-comment content constitutes the formula specification, which must evaluate to a Boolean value. The formula language supports propositional connectives (\verb|AND|, \verb|OR|, \verb|NOT|, \verb|IMPLIES|, \verb|IFF|), comparison operators (\verb|<|, \verb|>|, \verb|<=|, \verb|>=|, \verb|==|), arithmetic operators (\verb|+|, \verb|-|, \verb|*|, \verb|/|), and standard mathematical functions. Formula evaluation proceeds after all DP-cores have completed their tree decomposition traversals, substituting each variable with its acceptance value and each \verb|INV(x)| expression with the corresponding invariant value.

\paragraph{Numeric semantics.}
Numeric subexpressions are evaluated using double-precision floating-point arithmetic (IEEE-754) and the corresponding C++ standard-library functions. Division \verb|/| is real division. The function \verb|ln| denotes the natural logarithm and \verb|log| denotes the base-10 logarithm. Logical connectives treat \verb|0| as false and any non-zero value as true.

\paragraph{Parsing conventions.}
Whitespace is ignored. The formula header may be written as \verb|Formula| or \verb|FORMULA|. Boolean constants \verb|TRUE| and \verb|FALSE| are case-insensitive. All other operator/function keywords are case-sensitive as shown (e.g., \verb|INV|, \verb|IMPLIES|, \verb|IFF|). In addition to keyword forms, \verb|AND|/\verb|OR|/\verb|NOT| may also be written as \verb|&&|, \verb+|+, and \verb|!|, respectively.

The simplest formulas involve only Boolean combinations of variables:

\begin{code}
x := ChromaticNumber_AtMost(3)
y := ChromaticNumber_AtMost(4)
Formula
(NOT x) AND y
\end{code}

This formula accepts graphs that are $4$-colorable but not $3$-colorable. During evaluation, $\treewidzard$ processes the input graph through both DP-cores, obtains their Boolean values, and evaluates the formula through standard propositional logic.

Formulas may incorporate numerical invariants through comparison operators:

\begin{code}
x := IndependenceNumber()
Formula
INV(x) < 5
\end{code}

This formula accepts graphs whose independence number is strictly less than $5$. The DP-core \verb|IndependenceNumber| computes the independence number as its invariant, and the formula evaluates to true precisely when this computed value satisfies the inequality.

Complex formulas may combine Boolean logic, arithmetic operations on invariants, and mathematical functions:

\begin{code}
x := ChromaticNumber_AtMost(5)
y := IndependenceNumber()
z := VertexCount()
Formula
x IMPLIES (INV(y) > (((INV(z)/5) - 1)))
\end{code}

This formula expresses the implication: if a graph has chromatic number at most $5$, then its independence number must exceed $|V(G)|/5 - 1$. The evaluation proceeds by first computing the Boolean acceptance value $\chi(G) \leq 5$ for \verb|x| and the numeric invariant values $\alpha(G)$ and $|V(G)|$ for \verb|y| and \verb|z|, respectively, then evaluating the arithmetic expression $|V(G)|/5 - 1$, performing the comparison $\alpha(G) > |V(G)|/5 - 1$, and finally evaluating the implication.
The formula language supports a comprehensive set of operators and functions enabling expressive property specifications. Table~\ref{tab:formula_operators} provides a complete reference of supported constructs. Binary functions such as \verb|max|, \verb|min|, and \verb|pow| accept two arguments and return a single value. Unary functions apply standard mathematical operations to single arguments. All arithmetic operations follow standard precedence rules, with parentheses available for explicit grouping.

\begin{table}[H]
\centering
\renewcommand{\arraystretch}{1.4}
\begin{tabular}{| >{\centering\arraybackslash}m{4cm} | >{\centering\arraybackslash}m{10cm} |}
\hline
\textbf{Category} & \textbf{Operators / Functions} \\
\hline
Logical operators & \verb|AND|, \verb|OR|, \verb|NOT|, \verb|IMPLIES|, \verb|IFF| \\
\hline
Comparison operators & \verb|<|, \verb|>|, \verb|<=|, \verb|>=|, \verb|==| \\
\hline
Arithmetic operators & \verb|+|, \verb|-|, \verb|*|, \verb|/| \\
\hline
Binary functions & \verb|max|, \verb|min|, \verb|pow| \\
\hline
Unary functions & \verb|abs|, \verb|acos|, \verb|asin|, \verb|atan|, \verb|cos|, \verb|exp|, \verb|floor|, \verb|ln|, \verb|log|, \verb|sin|, \verb|sqrt|, \verb|tan| \\
\hline
Invariant access & \verb|INV(x)| \\
\hline
\end{tabular}
\caption{Operators and functions supported in property file formulas}
\label{tab:formula_operators}
\end{table}

\subsubsection{Comment Syntax}

Property files support line comments beginning with the character sequence \verb|//|. All content following \verb|//| on a line is ignored during parsing, enabling documentation and annotation. Comments may appear anywhere in the file: before variable declarations, interspersed with declarations, within the formula section, or at line endings following substantive content. The following example illustrates comment usage:

\begin{code}
// This property verifies a relationship between independence number and chromatic number
x := IndependenceNumber()
y := ChromaticNumber_AtMost(5)
// Formula Definition: independence number exceeds 5 if and only if graph is not 5-colorable
Formula
(INV(x) > 5) IFF (NOT y)
\end{code}

Comments serve multiple purposes: documenting the property's mathematical meaning, explaining formula logic, citing relevant theorem statements, and temporarily disabling portions of the specification during debugging. Judicious use of comments substantially improves property file maintainability and readability, particularly for complex formulas involving multiple DP-cores and intricate logical relationships.

\section{Unbounded ATP and Compositional Reasoning}
\label{appendix:unbounded_atp}

While the main text focuses on bounded DP-cores operating on graphs of bounded treewidth, $\treewidzard$ also supports specifications that involve \emph{unbounded} numeric invariants. In this regime, ATP exploration is not guaranteed to terminate, because even for a fixed width bound $k$ there exist graphs of width $\le k$ with arbitrarily many vertices.

\subsection{Unbounded Search Spaces}

To make this concrete, consider the invariant-returning DP-core \texttt{VertexCount()}, which computes the number of vertices $|V(G)|$. Since $|V(G)|$ is unbounded on the class of width-$\le k$ graphs, the induced ATP state space for formulas that depend on \texttt{VertexCount()} need not be finite, and the search may not reach a fixpoint under finite resource bounds. Nevertheless, such runs can still be useful for \emph{refutation}: counterexamples may appear early even when full exploration does not terminate.

\subsection{Bounding via Premise Pruning}

To enforce termination on a restricted domain, $\treewidzard$ supports premise pruning for implications $A \Rightarrow B$ where the premise $A$ is closed under subgraphs. A simple example is a vertex bound: the Boolean property $|V(G)|\le N$ is subgraph-closed (removing vertices cannot increase vertex count).

We can therefore bound an otherwise unbounded ATP exploration by using a premise of the form $(\mathrm{INV}(z)\le N)$ with $z:=\texttt{VertexCount()}$. For example:
\begin{code}
z := VertexCount()
Formula
( INV(z) <= 20 ) IMPLIES (1 == 1)
\end{code}
To enforce this bound operationally (i.e., to prune branches once \verb|INV(z) > 20|), the ATP run must be executed with the \texttt{-premise} flag. In this formulation, once the vertex bound fails on a partial graph, it remains false for all supergraphs along that branch, so the implication is already true downstream and the search can prune that branch. Operationally, bounding \verb|INV(z)| bounds the range of the \verb|VertexCount| invariant on explored partial graphs, which makes the induced global DP-state space finite (up to canonization) and restores termination/completeness on the premise-bounded domain.

\subsection{VertexCount: An Invariant DP-Core}
\label{subsec:vertex_count_core}

To demonstrate the implementation of invariant-computing DP-cores (as opposed to the decision DP-core \texttt{ChromaticNumber\_AtMost} and the optimization DP-core \texttt{IndependenceNumber}), we present the complete specification of the \verb|VertexCount| DP-core for graphs of bounded treewidth.

\subsubsection{Local Witness Structure}

The \verb|VertexCount| DP-core tracks a single piece of information: the number of vertices observed in the subgraph corresponding to the current node of the tree decomposition. The local witness structure is defined as:

\begin{code}[language=C++]
struct VertexCount_Witness : WitnessWrapper<VertexCount_Witness> {
    int vertexCount = 0;

    friend bool is_equal_implementation(
        const WitnessAlias &l, const WitnessAlias &r) {
        return l.vertexCount == r.vertexCount;
    }

    friend bool is_less_implementation(
        const WitnessAlias &l, const WitnessAlias &r) {
        return l.vertexCount < r.vertexCount;
    }

    WitnessAlias relabel_implementation(
        const std::map<unsigned, unsigned> &) const {
        auto relabeled = WitnessAlias();
        relabeled.vertexCount = vertexCount;
        return relabeled;
    }

    void hash(Hasher &h) const override {
        h << vertexCount;
    }

    void witness_info(std::ostream &os) const {
        os << vertexCount << '\n';
    }
};
\end{code}

The witness structure contains only the \verb|vertexCount| attribute, which is incremented each time a new vertex is introduced during the processing of the tree decomposition. Since vertex count is a simple numeric value, the relabeling operation (required for handling vertex permutations in graphs of bounded treewidth) acts as the identity function.

\subsubsection{DP-Core Structure and Metadata}

The \verb|VertexCount_Core| is neither a purely Boolean DP-core (like \verb|ChromaticNumber_AtMost|) nor a traditional optimization DP-core (like \verb|IndependenceNumber|). Instead, it computes a graph invariant, namely the total number of vertices, which is preserved under graph isomorphism.

\begin{code}[language=C++]
struct VertexCount_Core
    : CoreWrapper<VertexCount_Core, VertexCount_Witness,
                  WitnessSetTypeTwo> {

    static std::map<std::string, std::string> metadata() {
        return {
            {"CoreName", "VertexCount"},
            {"CoreType", "Min"},
            {"ParameterType", "None"},
        };
    }

    VertexCount_Core(const parameterType &parameters) {
        assert(parameters.empty());
        (void)parameters;
    }
};
\end{code}

The metadata specifies that this core requires no parameters (\verb|ParameterType| is \verb|None|), and the \verb|CoreType| is arbitrarily set to \verb|Min| since exactly one witness is produced for each graph of bounded treewidth, making the choice of minimization or maximization inconsequential.

\subsubsection{Transition Functions for Tree Decompositions}

The transition functions process nodes in the tree decomposition of graphs of bounded treewidth, maintaining the vertex count invariant at each step.

\paragraph{Leaf Initialization.}
Since the graph associated with a leaf node in the tree decomposition is empty, the witness set is initialized with a witness having \verb|vertexCount| set to zero:

\begin{code}[language=C++]
void initialize_leaf(WitnessSet &witnessSet) {
    witnessSet.insert(make_shared<WitnessAlias>());
}
\end{code}

\paragraph{Introducing a Vertex.}
When a new vertex is introduced, the vertex count must be incremented:

\begin{code}[language=C++]
void intro_v_implementation(unsigned, const Bag &,
                           const WitnessAlias &w,
                           WitnessSet &witnessSet) {
    auto wPrime = w.clone();
    ++wPrime->vertexCount;
    witnessSet.insert(wPrime);
}
\end{code}

\paragraph{Forgetting a Vertex.}
In tree decompositions, forgetting a vertex label does not remove the vertex from the graph—it merely removes the label from the active bag. Therefore, the vertex count remains unchanged:

\begin{code}[language=C++]
void forget_v_implementation(unsigned int, const Bag &,
                             const WitnessAlias &w,
                             WitnessSet &witnessSet) {
    witnessSet.insert(w.clone());
}
\end{code}

\paragraph{Introducing an Edge.}
Since introducing an edge does not affect the number of vertices, the witness is simply cloned:

\begin{code}[language=C++]
void intro_e_implementation(unsigned int, unsigned int,
                           const Bag &,
                           const WitnessAlias &w,
                           WitnessSet &witnessSet) {
    witnessSet.insert(w.clone());
}
\end{code}

\paragraph{Join Operation.}
The join operation combines two subgraphs in the tree decomposition that share a common bag $B$. Vertices in the bag $B$ are counted in both subtrees, so when computing the total vertex count, we must add the vertex counts from both witnesses and subtract the bag size to avoid double-counting:

\begin{code}[language=C++]
void join_implementation(const Bag &b, const WitnessAlias &w1,
                        const WitnessAlias &w2,
                        WitnessSet &witnessSet) {
    int joinedCount = w1.vertexCount + w2.vertexCount
                      - static_cast<int>(b.get_elements().size());
    auto wPrime = std::make_shared<WitnessAlias>();
    wPrime->vertexCount = joinedCount;
    witnessSet.insert(wPrime);
}
\end{code}

This calculation correctly handles the intersection of the two subgraphs at the shared bag, ensuring that the total vertex count matches $|V(G)|$ for the combined graph $G$ of bounded treewidth.

\subsubsection{Auxiliary Functions}

\paragraph{Final Witness Verification.}
All witnesses are considered valid, as each one correctly stores a vertex count:

\begin{code}[language=C++]
bool is_final_witness_implementation(const Bag &,
                                    const WitnessAlias &) {
    return true;
}
\end{code}

\paragraph{Clean Function.}
Witness cleaning removes redundant witnesses to improve performance. However, for \verb|VertexCount|, exactly one witness exists for each graph of bounded treewidth (no redundancy can occur), so the clean function is empty:

\begin{code}[language=C++]
void clean_implementation(WitnessSet &) {
}
\end{code}

\paragraph{Invariant Implementation.}
The invariant function returns the computed vertex count, making it available for use in compositional formulas via the \verb|INV()| operator:

\begin{code}[language=C++]
int inv_implementation(const Bag &, const WitnessAlias &w) {
    return static_cast<int>(w.vertexCount);
}
\end{code}

\section{Additional Counterexample Visualizations}
\label{appendix:additional_counterexamples}

This appendix provides visualizations of two small counterexamples generated by \treewidzard for
``algebraic degree bounds'' sanity checks.

\begin{figure}[H]
  \centering
  \includegraphics[width=0.92\linewidth]{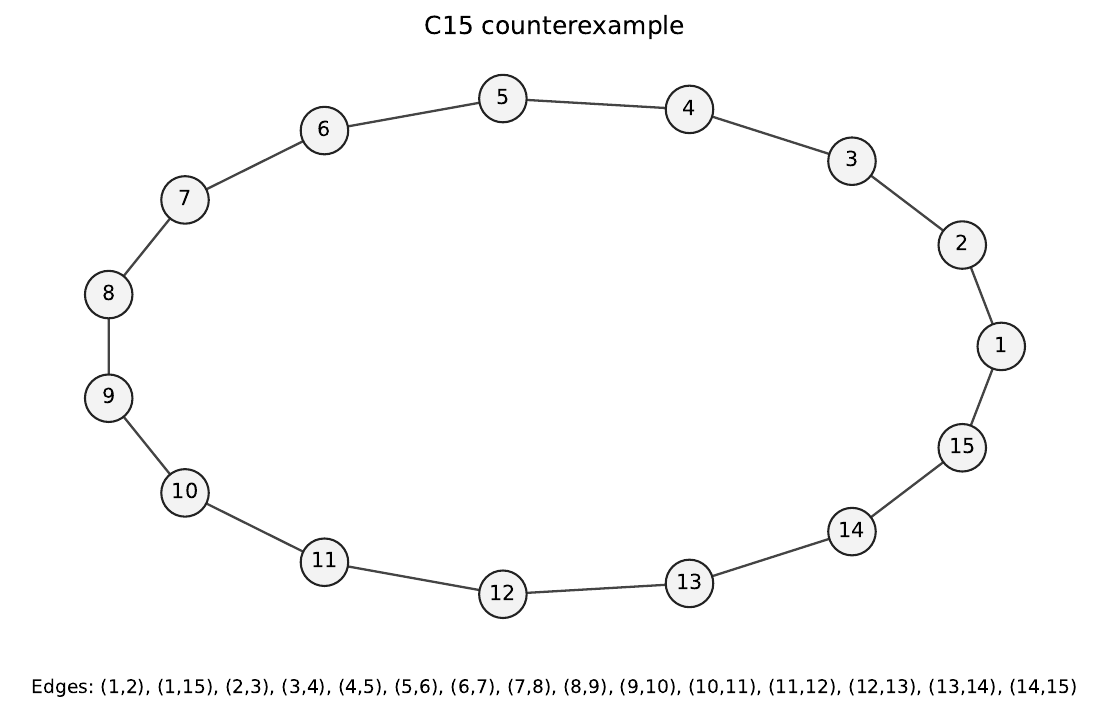}
  \caption{Counterexample on $15$ vertices refuting $\Delta \le 2 \Rightarrow 2\alpha \ge n$ (cycle $C_{15}$ with $\alpha=7$).}
  \label{fig:c15_counterexample}
\end{figure}

\begin{figure}[H]
  \centering
  \includegraphics[width=0.75\linewidth]{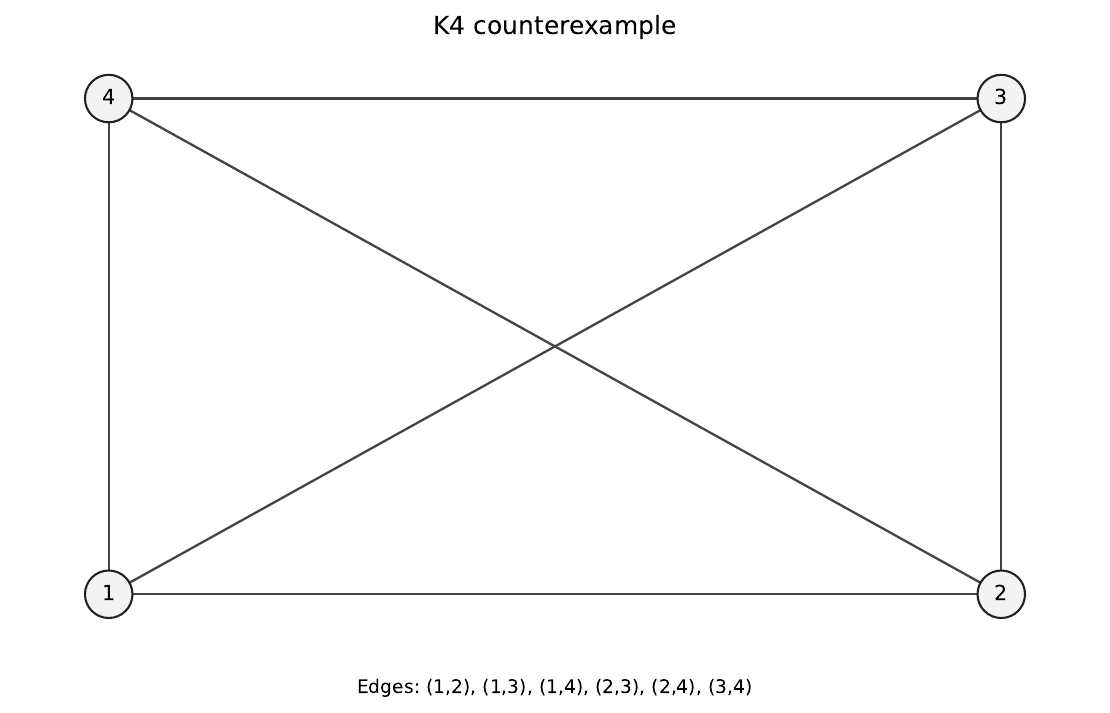}
  \caption{Counterexample refuting $\Delta = 3 \Rightarrow 2\alpha \ge n$ (clique $K_4$ with $\alpha=1$).}
  \label{fig:k4_counterexample}
\end{figure}

\end{document}